\documentclass[extra]{gji}
\usepackage{timet}
\usepackage{amsmath}
\usepackage{graphicx}
\usepackage{booktabs}
\usepackage{amssymb}
\usepackage{lineno}
\usepackage{placeins}

\renewcommand{\vec}{\mathbf}

\title[Dipole stability and zonal flows controlled by heat flux heterogeneities]
{Geomagnetic dipole stability and zonal flows controlled by mantle heat flux heterogeneities}
\author[T. Frasson, N. Schaeffer, H-C. Nataf, S. Labrosse]
{T. Frasson$^1$\thanks{thomas.frasson@univ-grenoble-alpes.fr}, N. Schaeffer$^1$, H-C. Nataf$^1$, S. Labrosse$^2$ \\
	$^1$ Univ. Grenoble Alpes, Univ. Savoie Mont Blanc, CNRS, IRD, Univ. Gustave Eiffel, ISTerre, 38000 Grenoble, France\\
	$^2$ ENS de Lyon, Universit\'e Lyon-1, LGL-TPE, 46 all\'ee d'Italie, 69007 Lyon.}

\begin{document}
	\label{firstpage}
	
	\maketitle

	\begin{summary}
		Palaeomagnetic evidence shows that the behaviour of the geodynamo has changed during geological times. These changes are visible through variations in the strength and stability of the magnetic dipole. Variations in the heat flux at the core-mantle boundary (CMB) due to mantle convection have been suggested as one possible mechanism capable of driving such a change of behaviour.
		This work aims at acquiring a more complete understanding of how lateral heterogeneities of the CMB heat flux affect the geodynamo while other relevant parameters are pushed towards realistic values.
		For this purpose, we ran geodynamo simulations with degree 1 and 2 spherical harmonic patterns of heat flux at the CMB. Several geodynamo models are used, ranging from standard numerical dynamos to more extreme parameters, including strong field cases and turbulent cases.
		We show that heat flux heterogeneities with amplitudes compatible with our knowledge of mantle convection history can favour multipolar dynamos. The multipolar transition is associated with a disruption of westward flows either through eastward thermal winds or through a loss of equatorial symmetry. 
		Strong field dynamo models are found to have larger westward flows and are less sensitive to heat flux heterogeneities.
		Furthermore, we find that the dipolar fraction of the magnetic field correlates with $M_{Za}^*=\dfrac{\Lambda_{Za}}{Rm_{Za}^2}$ where $\Lambda_{Za}$ is the zonal antisymmetric Elsasser number and $Rm_{Za}$ is the zonal antisymmetric magnetic Reynolds number. 
		Importantly, $M_{Za}^*$ estimated for the Earth's core is consistent with a reversing dipolar magnetic field.
		Within the range of $M_{Za}^*$ susceptible to reversals, breaking the equatorial symmetry or forcing eastward zonal flows through an equatorial cooling of the core consistently triggers reversals or a transition towards multipolar dynamos in our simulations.
		Our results support that time variations of heat-flux heterogeneities driven by mantle convection through Earth's history are capable of inducing the significant variations in the reversal frequency observed in the palaeomagnetic record.
	\end{summary}
	
	
	\section{Introduction}
	
	\subsection{Magnetic reversals and dipolar-multipolar transition in dynamo models}
	The magnetic field of the Earth is dominated by its dipole component. This magnetic dipole undergoes chaotic polarity reversals, with a reversal frequency that varies through geological times \citep{channell_geomagnetic_2013}. Reversals of the magnetic dipole have been reproduced in numerical dynamo models since the first attempts to simulate the geodynamo \citep{glatzmaier1995three}. These reversals are characterized by a transient decrease in the strength and stability of the magnetic dipole, which regrows and stabilizes with the opposite polarity \citep{olson2011complex}. Numerical models of reversing dynamos can be found at the transition between strongly dipolar dynamos and weakly dipolar (or multipolar) dynamos \citep{kutzner_stable_2002, olson_dipole_2006, wicht_theory_2010}. In the available set of dynamo models, this dipolar-multipolar transition seems to be controlled by the relative importance of inertia in the force balance \citep{christensen_scaling_2006, sreenivasan_azimuthal_2006}. The strength of inertial forces relative to Coriolis forces has notably been found to successfully grasp this transition in early dynamo models, larger inertia being associated with multipolar dynamos \citep{christensen_scaling_2006, olson_dipole_2006, christensen_dynamo_2010, wicht_theory_2010}.  
	In dynamo simulations operating in a strong magnetic field regime, in which the Lorentz force plays an important role as expected for the Earth, the relative strength of Coriolis forces and inertia does not accurately grasp the transition \citep{menu_magnetic_2020, tassin_geomagnetic_2021}.
	\citet{tassin_geomagnetic_2021} showed that the dipolar-multipolar transition can be more accurately described by the ratio of inertia to Lorentz forces, approximated by the kinetic over magnetic energy ratio. 
	According to this criterion, reversing dynamos should be obtained when the magnetic and kinetic energies within the core are of similar magnitudes. However, the Earth's dynamo seems to operate differently, with a magnetic energy several orders of magnitude higher than the kinetic energy. Though reversing dynamos producing a higher magnetic energy than kinetic energy on average have been obtained \citep{driscoll_effects_2009, nakagawa_combined_2022}, inertial forces are less than one order of magnitude weaker than Lorentz forces and tend to strengthen during reversals, suggesting that inertia is probably still important to drive reversals in these simulations. 
	
	\subsection{Mantle control on the behaviour of the magnetic dipole}
	Based on numerical dynamo models, different mechanisms have been proposed to explain the observed variations in the reversal frequency of the magnetic dipole. \citet{driscoll_effects_2009}, for example, suggest that variations in the control parameters of the geodynamo, such as the rotation rate or the power input, could explain regime changes. This mechanism requires the Earth to have been operating close to the inertia-driven dipolar-multipolar transition since the onset of dynamo action in the Earth's core. Using these assumptions, \citet{driscoll_polarity_2009} showed that variable reversal frequencies can be obtained in dynamo models with a simple core evolution model. 
	
	In this regard, the heat flux at the core-mantle boundary (CMB) can have an important role by modulating convective activity at the top of the core. Moreover, the CMB heat flux is expected to be strongly heterogeneous \citep{nakagawa_lateral_2008}. Large-scale heat flux heterogeneities have been shown to have a significant impact on the behaviour of the dynamo \citep{glatzmaier_role_1999, olson_time-averaged_2002, kutzner_simulated_2004, olson_geodynamo_2010, olson_magnetic_2014, amit_towards_2015, sahoo_dynamos_2016}. However, no clear trend has emerged. For instance, large heat flux heterogeneities have been shown to suppress dynamo action \citep{olson_time-averaged_2002}, while weak heterogeneities could on the contrary help sustain a dynamo \citep{sahoo_dynamos_2016}. 
	Interestingly, the reversal frequency has been shown to be very sensitive to heat flux heterogeneities. Various effects have been reported so far. Notably, large heat flux heterogeneities have been found to increase the reversal frequency, particularly for high equatorial heat fluxes \citep{kutzner_simulated_2004, olson_geodynamo_2010}.  
	Stabilizing effects of the magnetic dipole due to heat flux heterogeneities are also observed when polar cooling exceeds equatorial cooling \citep{glatzmaier_role_1999,kutzner_simulated_2004}.
	
	Recent models of the geodynamo showed that the present-day heat flux pattern, as inferred from seismic tomography, can explain the longitudinal structure of the observed magnetic field \citep{mound2023longitudinal}, and could trigger reversals from locally high heat flux regions \citep{terra2024regionally}. These findings suggest that the present-day heat flux pattern at the CMB plays an important role in the dynamics of the core. Though the present-day lateral heterogeneities of temperatures in the lower mantle are known to a certain extent \citep{trampert_probabilistic_2004, mosca_seismic_2012}, the past CMB heat flux is highly uncertain. On large scales, the present-day CMB heat flux is controlled by the positions of the observed Large Low Velocity Provinces (LLVPs) below Africa and the Pacific \citep{garnero2008structure}. These structures are interpreted as regions hotter than the surrounding mantle, thus locally extracting a low heat flux from the core \citep{nakagawa_lateral_2008}. Long-term stability of the observed LLVPs have been suggested based on records of hotspot eruption sites \citep{burke_plume_2008, torsvik2010diamonds, dziewonski_mantle_2010}, suggesting that the present-day heat flux pattern could be relevant for at least the past 300 Myr. However, mantle convection models show that basal mantle structures are pushed by subducting slabs and thus could have significantly been displaced through geological times \citep{zhang_heat_2011, flament2022assembly}. These variations in the CMB heat flux are expected to occur on timescales similar to the longest timescales of variations in the magnetic field, corresponding to changes in the reversal frequency \citep{biggin_possible_2012,hounslow_subduction_2018}. Understanding how the dynamo can react to a change in the CMB heat flux geometry is thus of importance to constrain the present-day geodynamo and its evolution through Earth's history.
	
	Mantle convection models can be used to have an estimate of the CMB heat flux heterogeneities in the past \citep{zhang_heat_2011}. Using self-consistent and plate-driven mantle convection models that reproduce $\sim 1$ Gyr of mantle convection, \citet{frasson2024impact} showed how episodes of true polar wander could significantly alter the CMB heat flux distribution on time scales that can be shorter than the convective time scale in the mantle. The strength of equatorial and polar cooling of the core is notably shown to depend on the way true polar wander rotates chemical structures at the base of the mantle. The mantle convection scenarios corrected from true polar wander presented in \citet{frasson2024impact} can thus be used to obtain estimates of heat flux heterogeneity amplitudes.
	
	\subsection{Aim of the study}
	
	Our current understanding of how heat flux heterogeneities affect the geodynamo is limited by the parameter regime reached in numerical models. In the last decade, new geodynamo simulations have reached a rapidly-rotating turbulent regime, in which the Lorentz force dominates inertial and viscous forces as expected for the Earth \citep{yadav2016approaching, schaeffer_turbulent_2017, aubert_spherical_2017}. Dynamos operating in this regime are computationally demanding, making parametric studies challenging. This challenge becomes even more important when rare events such as reversals are targeted. For this reason, previous studies used dynamo models that are only moderately turbulent. Inertia and viscous forces have been shown to be non-negligible in these types of dynamos \citep{soderlund2012influence, dormy2016strong}, which are therefore not representative of the Earth's core.
	
	In this study, we aim at exploring the effect of heat flux heterogeneities on dynamo models that extend from standard dynamo regimes towards a more Earth-like rapidly-rotating turbulent regime. We first study the effect of large scale heat flux patterns (spherical harmonic degrees 1 and 2) on a standard dynamo reference case. We then focus on the effect of polar cooling and equatorial cooling on more turbulent dynamos, exploring more Earth-like parameter regimes. We notably focus on the reversing behaviour and the dipolar-multipolar transition triggered by heat flux heterogeneities. 
	
	\section{Methods}
	\label{s:m&m}
	
	\subsection{Numerical model of the geodynamo}
	
	The Earth's liquid outer core is modelled as a spherical shell of inner radius $r_i$ and outer radius $r_o$. The thickness of the shell is defined as $D = r_o - r_i$. The aspect ratio is chosen to match the present-day Earth by fixing $r_i/r_o = 0.35$. The shell spins along the $z$ axis at a rate $\Omega$. The acceleration of gravity is defined as $\mathbf{g} = -\textsl{g}_o\frac{r}{r_o}\vec{e_r}$ with $\textsl{g}_o$ the acceleration of gravity at the core surface and $\vec{e_r}$ the unit vector in the radial direction. We use the Boussinesq approximation in which the electrically conducting liquid contained in the shell is treated as an incompressible fluid of density $\rho_0$ and viscosity $\nu$. The electrical properties of the fluid are defined by its magnetic permeability $\mu_0$ and its magnetic diffusivity $\eta$. Convection in the Earth's outer core is the result of both thermal and compositional heterogeneities, which translate into density heterogeneities. Here, we use the codensity approximation defined in \citet{braginsky_equations_1995}, which is exact when both thermal and compositional diffusivities are equal, and when the same boundary conditions apply to temperature and composition fields. The thermal and compositional sources of buoyancy are then combined into one single variable called codensity, with diffusivity $\kappa$. If we call $T$ the temperature field, $\chi$ the concentration in light element, and $\rho_0$ the reference density, the codensity is defined as $C = \rho_0 (\alpha T + \beta \chi$) with $\alpha = -\dfrac{1}{\rho_o}\dfrac{\partial \rho}{\partial T}$ and $\beta = -\dfrac{1}{\rho_o}\dfrac{\partial \rho}{\partial \chi}$ the thermal and compositional expansion coefficients respectively. Convection in the shell is driven by imposed buoyancy fluxes at the inner and outer boundaries. The buoyancy fluxes per unit surface (in kg.m$^{-2}$.s$^{-1}$) at the inner and outer boundaries are called $q_i$ and $q_o$ respectively, and are defined by 
	\begin{linenomath}
		\begin{align}
			q_i &= -\kappa \ \partial_r C (r_i),\\
			q_o &= -\kappa \ \partial_r C (r_o).
		\end{align}
	\end{linenomath}
	The buoyancy flux entering the core at the inner boundary $F_i$ and the buoyancy flux leaving the core at the outer boundary $F_o$ can then be obtained by integration over the boundaries:
	\begin{linenomath}
		\begin{align}
			F_i &= \int_{S_i} q_i\ dS_i,\\
			F_o &= \int_{S_o} q_o\ dS_o.
		\end{align}
	\end{linenomath}
	where $S_i$ and $S_o$ are the inner and outer boundaries respectively. 
	
	The geodynamo equations consists in the coupling of Navier-Stokes and induction equations. We solve this set of equations in their dimensionless form using the Boussinesq approximation. 
	The thickness of the shell $D$ is used as the length scale. The time is scaled by the viscous timescale $\tau_{\nu} = \frac{D^2}{\nu}$ and the magnetic field by $\sqrt{\rho_0 \Omega \eta \mu_0}$. The codensity is scaled by $\frac{F_i}{4\pi D \nu}$. 
	The codensity field $C$ is decomposed into a base radially-dependent profile $C_0$ and a perturbation $C'$, such that $C = C_0 + C'$. 
	Calling $\vec{u}$ the velocity field, $P$ the pressure, and $\vec{B}$ the magnetic field, the set of dimensionless equations is as follow:
	\begin{linenomath}
		\begin{align}
			\begin{split}
				\partial_t \vec{u} + & (\nabla \times \vec{u}) \times \vec{u} + \dfrac{2}{E} \vec{e_z} \times \vec{u} = -\nabla P + \nabla^2 \vec{u} \\
				& + \dfrac{1}{Pm\ E} (\nabla \times \vec{B}) \times \vec{B} + \dfrac{Ra}{Pr} C' \dfrac{r}{r_o} \vec{e_r},
			\end{split}
		\end{align}
	\end{linenomath}
	\begin{linenomath}
		\begin{align}
			\partial_t C' + \vec{u} \cdot \nabla (C' + C_0) &= \dfrac{1}{Pr} \nabla^2 (C' + C_0) + S_c,\\
			\label{codensity}
			\partial_t \vec{B} &= \nabla \times (\vec{u} \times \vec{B}) + \dfrac{1}{Pm} \nabla^2 \vec{B},\\
			\nabla \cdot \vec{u} &= 0\\
			\nabla \cdot \vec{B} &= 0
		\end{align}
	\end{linenomath}
	where $S_c$ is a sink term to compensate for the imbalance between $F_i$ and $F_o$. The dimensionless parameters $E$, $Pm$, $Pr$ and $Ra$ are respectively the Ekman number, the magnetic Prandtl number, the Prandtl number and the Rayleigh number. They are defined using the physical parameters of the system by 
	\begin{linenomath}
		\begin{align}
			E &= \dfrac{\nu}{\Omega D^{2}},\\
			Pm &= \dfrac{\nu}{\eta},\\
			Pr &= \dfrac{\nu}{\kappa},\\
			Ra &= \dfrac{\textsl{g}_oF_iD^{2}}{4 \pi \kappa \nu^{2} \rho_0}.
		\end{align}
	\end{linenomath}
	
	No slip conditions are applied at both boundaries for the velocity. The outer boundary is electrically insulating in all our simulations. The inner core is treated as a conducting sphere in all but one of our dynamo models, with a conductivity of the inner core equal to that of the outer core. An insulating inner core is used for the most numerically demanding simulations. An heterogeneous flux condition at the outer boundary can be imposed. This flux pattern is described by an amplitude $\delta q_o$ and a pattern $p(\theta, \phi)$, which depends on the colatitude $\theta$ and the longitude $\phi$. We impose that the heterogeneous flux has a zero space average and set $F_o=0$. 
	
	In this codensity formalism, the flux imposed at either of the boundaries is a non-distinguishable combination of a heat flux and a light-element flux. This buoyancy flux is expected to be of thermal origin at the top of the core, with a null compositional flux, and mostly of compositional origin at the inner boundary. In the following, we will thus refer to the flux at the top of the core as a heat flux. As we use the Boussinesq framework, the imposed fluxes at the boundaries correspond to super-isentropic fluxes, and the isentropic flux would have to be added to obtain the total fluxes entering and leaving the core. This means notably that a negative flux at the top of the core tends to act against convection by stratifying the fluid. 
	
	The equations are solved using the \textit{XSHELLS} code benchmarked in \citet{matsui2016performance} and already used in previous studies \citep{schaeffer_turbulent_2017, guervilly2019turbulent}. Finite differences are used in the radial direction, while the fields are expanded in spherical harmonics in the horizontal directions. The spherical harmonics transforms are performed using \textit{SHTns} \citep{Schaeffer2013efficient}. We use hyperdiffusivity to damp small scales and stabilize the simulations in some cases as done in \citet{guervilly2019turbulent} (see appendix \ref{app1} for more details).

	\subsection{Output parameters of interest}
	\label{outputs}
	The simulation outputs are given in the following as a function of time scaled by the magnetic dipole diffusion timescale rather than the viscous timescale used to solve the equations. The dipole diffusion timescale is defined by 
	\begin{linenomath}
		\begin{equation}
			\tau_{\eta}=\dfrac{1}{\pi^2} \dfrac{r_o^2}{\eta}
		\end{equation} 
	\end{linenomath}
	and can be estimated to be approximately $40$ kyr for the Earth using $\eta = 1$ m$^2$ s$^{-1}$ \citep{olson_801_2015}, though the exact value of $\eta$ is prone to some debate \citep{pozzo2014thermal, holdenried2022evidence}.
	The conversion between the simulation time $t$ and the output time $t_{\eta}$ is then simply $t_{\eta} = t \,  \dfrac{\tau_{\nu}}{\tau_{\eta}}$.
	
	Several output parameters are used to compare our simulations between each other and to the Earth. The strength of the magnetic dipole is quantified by the dipolar fraction $f_{dip}$ defined at the CMB as 
	\begin{linenomath}
		\begin{equation}
			f_{dip}(t) = \sqrt{\dfrac{\sum\limits_{m=0}^{1}(2-\delta_{m0})|b_r(l=1,m,t)|^2}{\sum\limits_{l,m\ge0}(2-\delta_{m0})|b_r(l,m,t)|^2}}
		\end{equation}
	\end{linenomath}
	where $l$ and $m$ are the spherical harmonic degree and order, $b_r(l,m)$ is the spherical harmonic coefficient of degree $l$ and order $m$ of the radial component of the magnetic field at the CMB, 
	and $\delta_{lm}$ is the Kronecker delta. 
	In order to compare our results to the Earth's magnetic field, we also compute an alternative dipolar fraction based only on the degrees $l\le 12$ of the magnetic field:
	\begin{linenomath}
		\begin{equation}
			f_{dip}^{\dagger}(t) = \sqrt{\dfrac{\sum\limits_{m=0}^{1}(2-\delta_{m0})b(l=1,m,t)^2}{\sum\limits_{l,m\ge 0}(2-\delta_{m0})b(l\le12,m,t)^2}}.
		\end{equation}
	\end{linenomath}
	This definition of the dipolar fraction enables to compare with measurements of the magnetic field that are limited to a maximum spherical harmonic degrees $l<13$ and is commonly used as outputs in dynamo simulations \citep{christensen_scaling_2006}. 
	The dimensionless kinetic and magnetic energy integrated over the whole spherical shell are respectively defined as 
	\begin{linenomath}
		\begin{align}
			E_u(t) &= \dfrac{1}{2} \int_{V_f} \vec{u}(t)^2 dV,\\
			E_b(t) &= \dfrac{1}{2} \int_{V_{c}} \left(\dfrac{\vec{B}(t)}{\sqrt{EPm}}\right)^2 dV
		\end{align}
	\end{linenomath}
	where $V_f$ is the volume of the fluid outer core and $V_c$ the volume of the conducting shell (corresponding to the full core if the inner core is conducting, or only the fluid outer core if the inner core is insulating). The ratio of these two quantities is called $M$ in the following, and is defined as
	\begin{linenomath}
		\begin{equation}
			M=E_b/E_u.
		\end{equation}
	\end{linenomath}
	This ratio is used as a proxy of the ratio between the Lorentz force and the inertial forces by \citet{tassin_geomagnetic_2021}, and is found to discriminate between dipole-dominated dynamos and multipolar dynamos, with a transition around $M \simeq 1$. The amplitudes of the velocity field and of the magnetic field are quantified by the magnetic Reynolds number $Rm=\frac{U D}{\eta}$ and the Elsasser number $\Lambda=\frac{B^2}{\rho_0\Omega \mu_0 \eta}$ where $U$ and $B$ are the flow and magnetic field RMS amplitudes respectively. Using the dimensionless outputs of the simulation, those numbers are computed as
	\begin{linenomath}
		\begin{align}
			Rm(t) &= Pm\ \sqrt{\dfrac{2 E_u(t)}{V_f}},\\
			\Lambda(t) &= \dfrac{2Pm\ E}{V_f} E_b(t).
		\end{align}
	\end{linenomath}
	The strength of inertial force relative to Coriolis force can be approximated using the local Rossby number \citep{christensen_scaling_2006}
	\begin{linenomath}
		\begin{equation}
			\label{Rol}
			Ro_l(t) = \dfrac{E\ l_c(t)}{\pi}\ \sqrt{\dfrac{2 E_u(t)}{V_f}} 
		\end{equation}
	\end{linenomath}
	where $l_c$ is the characteristic spherical harmonic degree of the flow. This characteristic degree is defined by 
	\begin{linenomath}
		\begin{equation}
			l_c(t) = \dfrac{\sum_l l E_{u,l}(t)}{E_u(t)}
		\end{equation}
	\end{linenomath}
	with $E_{u,l}$ the kinetic energy at degree $l$. Finally, we quantify the stability of the magnetic dipole by computing the average period spent by the dipole in a given hemisphere. This period, called $T_{rev}$, is computed for the reversing dynamos as
	\begin{linenomath}
		\begin{equation}
			\label{Trev}
			T_{rev} = \dfrac{\Delta t_{sim}}{N_{cross}}
		\end{equation}
	\end{linenomath}
	where $\Delta t_{sim}$ is the duration of the equilibrated simulation and $N_{cross}$ is the number of times the magnetic dipole crosses the equator. We note that the value of $T_{rev}$ is not very robust when $N_{cross}$ is small, which can happen for short simulations or when the reversal rate is low.
	
	Because the simulations start from an initial condition that is different from their statistically steady state, the transient evolution from the initial condition has to be removed before computing statistics on the diagnostics. We ensure that this transitional state is not affecting the results by removing the evolution corresponding to the first two dipole diffusion times.

	\subsection{Explored parameter space}
	\label{s:parameter}
	
	\subsubsection{Reference dynamo models}
	
\begin{figure}
		\centering
		\includegraphics[width=\linewidth]{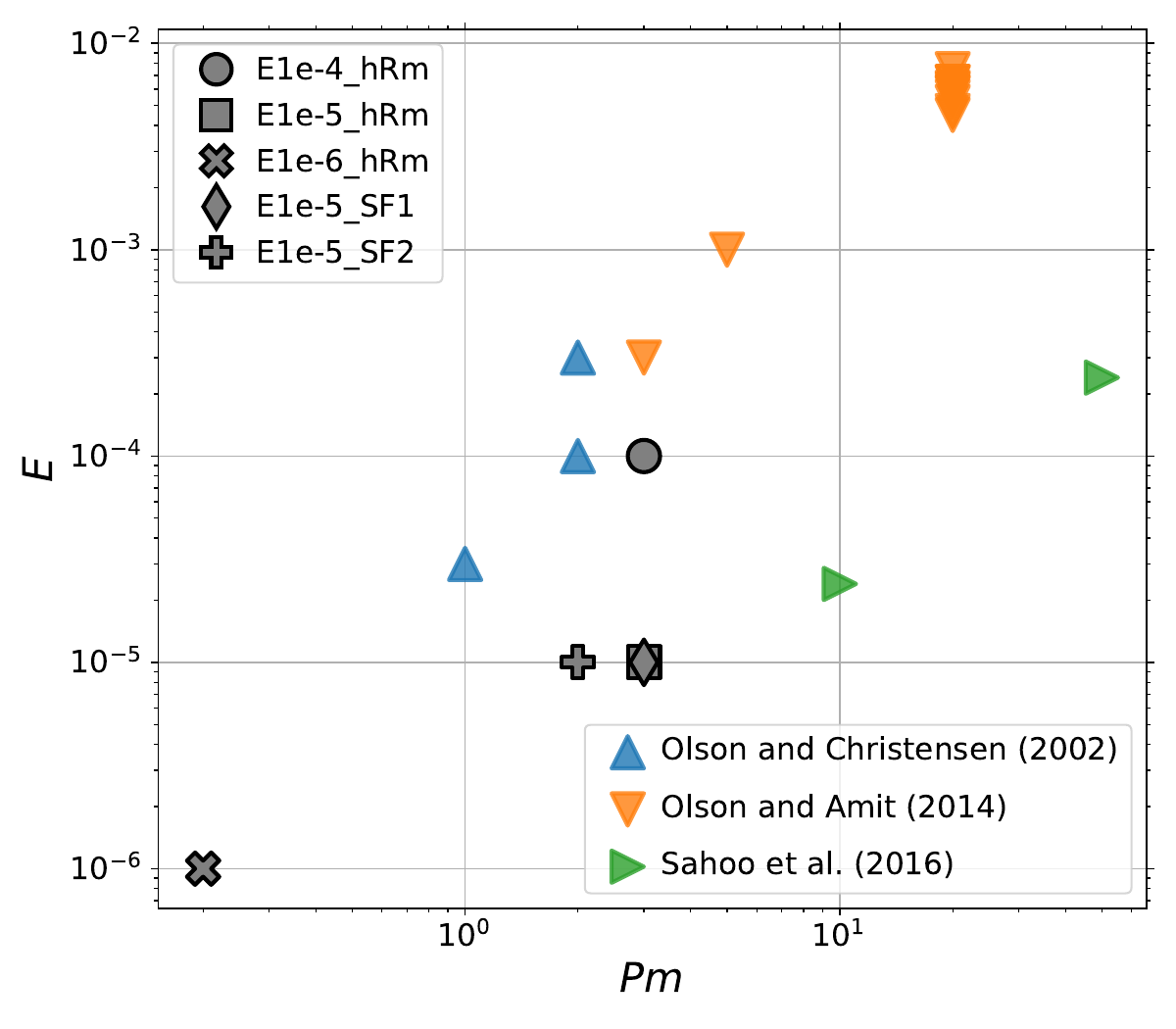}
		\caption{Ekman number $E$ and magnetic Prandtl number $Pm$, for the 5 reference dynamos E1e-4\_hRm, E1e-5\_hRm, E1e-6\_hRm, E1e-5\_SF1, and E1e-5\_SF2 compared to previous studies focusing on the effect of large scale heterogeneous heat flux \citep{olson_time-averaged_2002, olson_magnetic_2014, sahoo_dynamos_2016}. All the dynamo models are obtained for $Pr=1$. The Rayleigh number of the dynamo models are $1.5\times 10^8$, $5\times 10^9$, $9.6\times 10^{11}$,  $5\times 10^8$, and $1.5\times 10^8$  for E1e-4\_hRm, E1e-5\_hRm, E1e-6\_hRm, E1e-5\_SF1, and E1e-5\_SF2 respectively. The expected values for the Earth are $E\simeq10^{-15}$ and $Pm\simeq10^{-6}$.}
		\label{Ra_E_references}
	\end{figure}

	We select 5 reference dynamo cases with homogeneous heat flux conditions at the top of the core, to which heterogeneous heat flux patterns are applied.
	In Fig. \ref{Ra_E_references} these dynamos are compared in the $E-Pm$ parameter space with previous studies that focused on the effect of heterogeneous heat flux on the dynamo. Our reference case E1e-4\_hRm is comparable to dynamo models already used in previous studies, with $E=10^{-4}$ and $Pm=3$. Our cases E1e-5\_hRm, E1e-5\_SF1 and E1e-5\_SF2 have similar magnetic Prandtl number, but have a lower viscosity than previous studies. Finally, case E1e-6\_hRm is getting closer to the parameter regime of the Earth's core with both a lower Ekman number and a lower magnetic Prandtl number ($E=10^{-6}$, $Pm=0.2$).
	
	The values of the parameters introduced in section \ref{outputs} can be estimated to a certain extent for the Earth. Values for the magnetic Reynolds number and the Elsasser number can be estimated to $Rm=400-4000$ and $\Lambda=5-50$ \citep{aubert_spherical_2017}. Similarly the ratio of magnetic to kinetic energy is likely in the range $M=10^3-10^4$ \citep{tassin_geomagnetic_2021}. For the dipolar fraction of the magnetic field, we will here take the range $f_{dip}^{\dagger}=0.35-0.75$ suggested by \citet{davies2022dynamo}.
	Table \ref{Rm_fdip_refs} compares the time-averaged values of the dipolar fraction, the energy ratio, the Elsasser number, and the magnetic Reynolds number of the reference cases with the Earth estimates. All the reference cases have energy ratios that are far too small compared to the expectation for the Earth. Cases E1e-5\_SF1 and E1e-5\_SF2 nevertheless have $\overline{M}\gg10$, the overline standing for time average. For this reason, these two cases are called ``SF'' for ``Strong Field'' \citep{schwaiger2021relating}. The magnetic over kinetic energy ratio is of order 1 or larger for the three other cases. However, these three cases have magnetic Reynolds numbers that match the values expected for the Earth, while the two strong field dynamos have values of $Rm$ that are too low or just at the limit of the Earth's range. We thus call these three cases ``hRm'' for ``high $Rm$''. Cases E1e-4\_hRm, E1e-5\_hRm and E1e-5\_SF1 have dipolar fractions within the Earth's range, while E1e-5\_SF2 and E1e-6\_hRm have larger dipolar fractions. All the cases have Earth-like Elsasser numbers. It can be noted that the strong field cases have low magnetic Reynolds numbers, while the high $Rm$ dynamo cases have more comparable magnetic and kinetic energies with $M\gtrsim1$. The inability to meet both criteria for a single dynamo model is a consequence of the distance between the parameter regime of dynamo models and the parameter regime of the Earth's core \citep{aubert_spherical_2017, nataf2024dynamic}. In this study, we consider dynamo models that satisfy one criterion or the other in order to examine whether these two types of dynamo models behave differently in the presence of heat flux heterogeneities. This choice enables us to study the effect of heat flux heterogeneities on models close to the dipolar multipolar transition (high $Rm$ models) and on models away from the transition (strong field models).
	
	\begin{table}
		\centering
		\caption{Characteristics of the reference dynamos compared to estimated values for the Earth. $^a$\citet{davies2022dynamo}; $^b$\citet{tassin_geomagnetic_2021}; $^c$\citet{aubert_spherical_2017}.}
		\begin{tabular}{lccccc}
			Name & $\overline{f_{dip}}$ & $\overline{f_{dip}^{\dagger}}$ & $\overline{M}$ & $\overline{\Lambda}$ & $\overline{Rm}$\\
			\midrule
			E1e-4\_hRm & 0.45 & 0.58 & 0.95 & 13.1 & 649\\
			E1e-5\_hRm & 0.38 & 0.53 & 3.49 & 16.8 & 987\\
			E1e-6\_hRm & 0.72 & 0.80 & 3.53 & 5.8 & 575\\
			E1e-5\_SF1 & 0.61 & 0.70 & 32.12 & 16.9 & 399\\
			E1e-5\_SF2 & 0.83 & 0.79 & 53.50 & 8.13 & 214\\
			Earth & -- & 0.35--0.75$^a$ & $10^3$--$10^4$$^b$ & 5--50$^c$ & 400--4000$^c$
		\end{tabular}
		\label{Rm_fdip_refs}
	\end{table}
	
	\subsubsection{Heat flux patterns}
	
\begin{figure*}
		\centering
		\includegraphics[width=\linewidth]{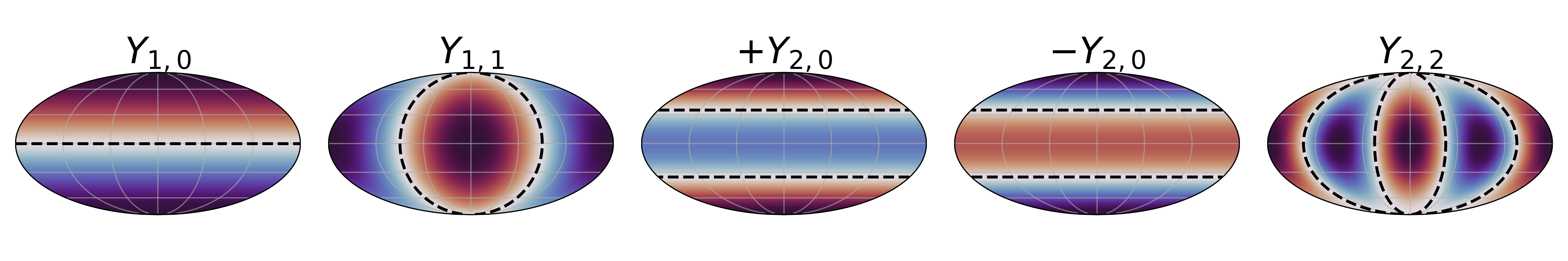}
		\caption{Heterogeneous heat flux patterns $p(\theta,\phi)$ used in the dynamo simulations. The positive and negative heat flux areas are shown in red and blue respectively. The dashed lines show the positions where the patterns are equal to zero. The patterns are shown in a Mollweide projection.}
		\label{patterns}
	\end{figure*}

	We study the effect of heterogeneous heat flux patterns at the top of the core applied to the reference dynamo cases described above. We focus on large scale heterogeneity patterns with spherical harmonics of degree 1 or 2. The patterns are shown in Fig. \ref{patterns}. Note that, in this figure, areas shown in red and blue correspond to a positive and negative heat flux respectively, i.e an increased and decreased heat flux from the core to the mantle respectively. Pattern $Y_{1,0}$ corresponds to a positive heat flux in the northern hemisphere and a negative heat flux in the southern hemisphere. Pattern $Y_{1,1}$ is the same as $Y_{1,0}$, but tilted by 90° in latitude. $+Y_{2,0}$ imposes a positive heat flux at the poles and a negative heat flux at the equator. $-Y_{2,0}$ is the opposite of $+Y_{2,0}$, imposing a negative heat flux at the poles and a positive heat flux at the equator. Finally, $Y_{2,2}$ imposes two antipodal heat flux highs at the equator, surrounded by two antipodal heat flux lows. Negative amplitudes for the $Y_{1,1}$ and $Y_{2,2}$ patterns only implies a rotation in longitude of the patterns that is irrelevant for the dynamo. Neither does a sign change of the $Y_{1,0}$ pattern affect the behaviour of the dynamo due to the symmetry of the dynamo equations. These heat flux patterns directly affect the flow through the generation of temperature gradients at the top of the core. The $Y_{1,1}$ and $Y_{2,2}$ patterns have been shown to generate an equatorial circulation with localized downwellings and upwellings offset from the high and low heat flux patches \citep{zhang1992convection, dietrich2016core}, which is verified in our simulations. The $Y_{1,0}$ and $Y_{2,0}$ patterns generate thermal winds that are equatorially antisymmetric for $Y_{1,0}$ and dominantly westward/eastward for $+Y_{2,0}$ and $-Y_{2,0}$ respectively \citep{amit2011influence,dietrich2017reversal}. 
	The imposed heat flux at the CMB is given by the product of an amplitude $\delta q_o$ and a pattern $p$,
	\begin{linenomath}
		\begin{equation}
			q_o(\theta, \phi) = \delta q_o\ p(\theta, \phi).
		\end{equation}
	\end{linenomath}
	The normalisation of the pattern $p$ ensures that $2\delta q_o = max(q_o) - min(q_o)$. We choose to scale the heterogeneity amplitude using the buoyancy flux at the inner boundary by defining a dimensionless amplitude $\delta q_o^*=\delta q_o/q_i$. Our scaling differs from most previous studies, where heat flux heterogeneities have been more commonly scaled by the mean outer boundary flux \citep{olson_geodynamo_2010, olson_magnetic_2014, mound2023longitudinal, terra2024regionally, dannberg2024changes}, but it has the advantage of being well defined even when $F_o=0$. It is also relevant for the Earth, for which the main forcing is thought to occur at the inner core boundary in the form of a compositional flux \citep{loper1978}, while the radial temperature profile could be close to isentropic ($F_o\simeq0$). The same scaling was used by \citet{aubert2013bottom} and \citet{nakagawa_combined_2022}. The heterogeneity amplitude normalized this way can be expressed for the Earth as 
	\begin{linenomath}
		\begin{equation}
			\label{def_dqo}
			\delta q_o^* = \dfrac{\alpha\delta q_o^{th}}{c_p \dot{r}_i\left(\Delta \rho + \dfrac{\alpha \rho_c L}{c_p}\right)}
		\end{equation}
	\end{linenomath}
	where $2 \delta q_o^{th}$ is the peak-to-peak amplitude of the CMB heat flux, $c_p$ is the heat capacity of the liquid iron, $\dot{r}_i$ is the growth rate of the inner core, $\Delta \rho$ is the compositional density jump at the inner core boundary, $\rho_c$ is the mean density of the core, and $L$ is the latent heat of liquid iron crystallisation. Table \ref{dqo_table} shows estimates of the maximum values of $\delta q_o^*$ using mantle convection simulations corrected for true polar wander by \citet{frasson2024impact}. The amplitude of the total heat flux pattern heterogeneities reach up to $\delta q_o^*=10.3\%$. The heterogeneities for a given spherical harmonic coefficient are lower, but can be as high as $\delta q_o^*=3.4\%$ for $Y_{2,2}$.
	Note that we impose $F_o=0$ and use the Boussinesq approximation in our dynamo models, so that the average heat flux extracted by the mantle is identified with the isentropic heat flux and the flux of light elements between the core and the mantle is null. The heterogeneous flux will thus be stabilizing in areas where $q_o(\theta,\phi)<0$ (i.e. lower than the isentropic flux). 
	
	All five patterns have been applied to the reference case E1e-4\_hRm. The four other cases require much more computing resources due to their lower Ekman numbers. We thus focus on the $\pm Y_{2,0}$ patterns for cases E1e-5\_hRm, E1e-6\_hRm, E1e-5\_SF1, and E1e-5\_SF2.
	
	\begin{table*}
		\caption{Maximal heterogeneity amplitude in the mantle convection simulations corrected for true polar wander by \cite{frasson2024impact}. The values below label $\delta {q_{o}^{*}}_{max}$ give the maximum peak-to-peak amplitude in the total heat flux patterns for the different cases. The values below $\delta q_o^{*}\left(Y_{l,m}\right)_{max}$ give the maximum peak-to-peak amplitude of the degree $l$ and order $m$ of the CMB heat flux time series in the different cases. For the $Y_{2,0}$ pattern, the first value gives the minimum negative amplitude (corresponding to $-Y_{2,0}$) and the second gives the maximum positive amplitude (corresponding to $+Y_{2,0}$). 
			In each column, we emphasize with bold font the highest value reached among the mantle convection cases (all defined in \citet{frasson2024impact}).}
		\begin{tabular}{ccccccc}
			Case&$\delta {q_{o}^{*}}_{max}$&$\delta q_o^{*}\left(Y_{1,0}\right)_{max}$&$\delta q_o^{*}\left(Y_{1,1}\right)_{max}$&$\delta q_o^{*}\left(-Y_{2,0}\right)_{max}$&$\delta q_o^{*}\left(+Y_{2,0}\right)_{max}$&$\delta q_o^{*}\left(Y_{2,2}\right)_{max}$\\
			\hline
			MF1&\textbf{10.3\%}&1.6\%&2.4\%&\textbf{2.9}\%& 2.6\%&3.2\\
			MF2&\textbf{10.3\%}&\textbf{1.7\%}&2.3\%&0\%&\textbf{2.7\%}&3.1\%\\
			MF$^*$&\textbf{10.3\%}&1.2\%&\textbf{2.6\%}&2.4\%&1.8\%&\textbf{3.4}\%\\
			MC1&7.1\%&0.9\%&1.3\%&1.3\%&0\%&1.0\%\\
		\end{tabular}
		\label{dqo_table}
	\end{table*}
	
	\section{Results}
	\label{s:results}

	\subsection{Reference homogeneous dynamos}
	\label{s:homogeneous}
	
	We shortly describe here the reference dynamo simulations with homogeneous heat flux. The five reference dynamos cover a wide range of parameters, with Ekman numbers between $10^{-4}$ and $10^{-6}$ and magnetic Prandtl numbers between 3 and 0.2 (Fig. \ref{Ra_E_references}). The Rayleigh number is also significantly varied among the reference cases.
	
\begin{figure*}
		\centering
		\includegraphics[width=\linewidth]{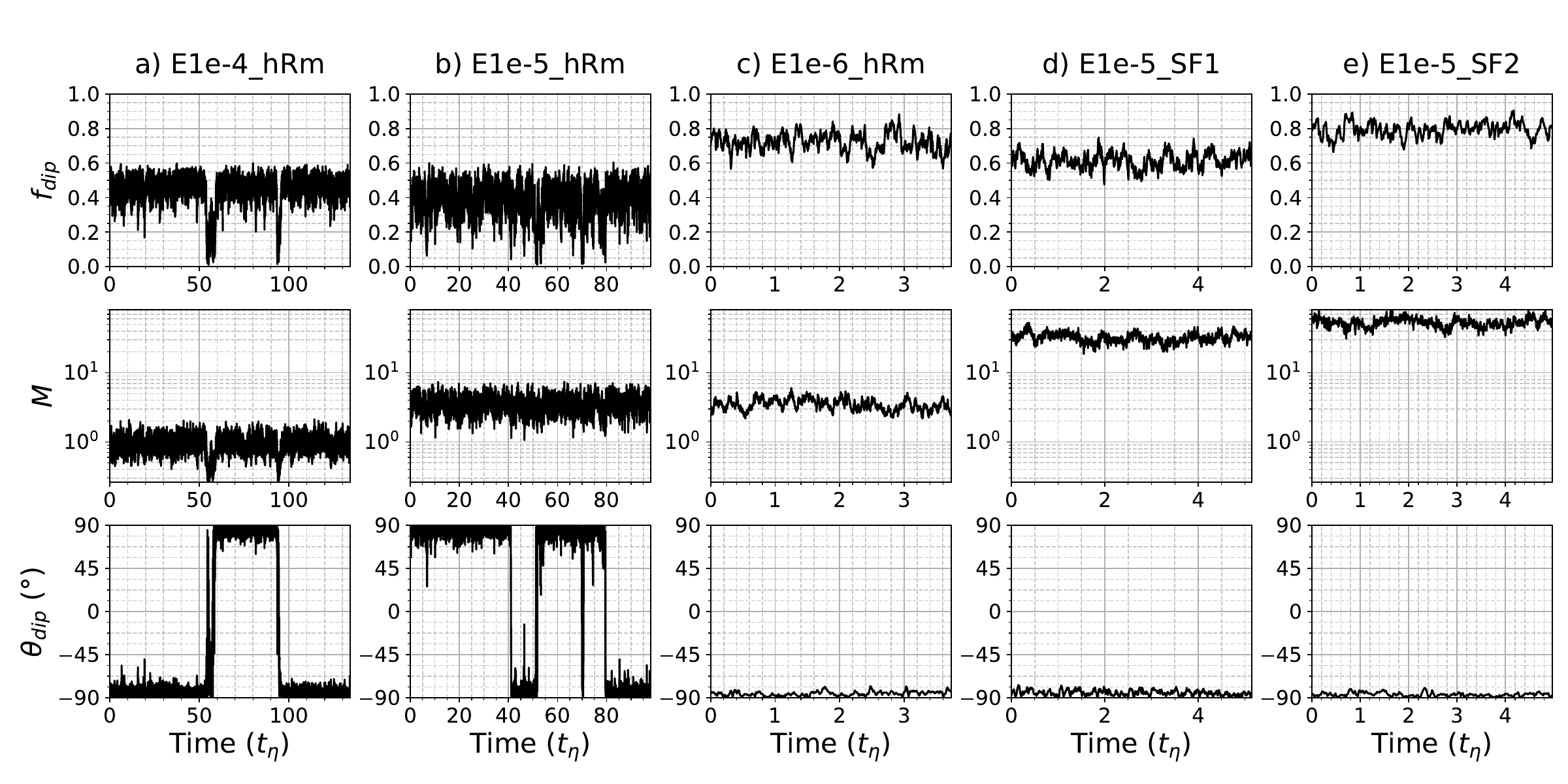}
		\caption{Time series of the dipolar fraction ($f_{dip}$), the ratio of magnetic to kinetic energy ($M$) and of the dipole latitude for the five reference geodynamo simulations considered in this study. Note that the horizontal scale for time is different for each case. The two first dipole diffusion time of the simulations have been removed for E1e-6\_hRm, E1e-5\_SF1 and E1e-5\_SF2. The simulations for E1e-4\_hRm and E1e-5\_hRm have been started from an already converged state.}
		\label{reference}
	\end{figure*}

	In Fig. \ref{reference} are shown the time evolution of the dipolar fraction $f_{dip}$, the ratio of magnetic to kinetic energy $M$, and the dipole latitude (angle between the equator and the magnetic dipole) $\theta_{dip}$. All the reference cases have strong dipoles with $\overline{f_{dip}} \ge 0.39$. The E1e-4\_hRm reference case displays magnetic reversals. The ratio of magnetic to kinetic energy is around unity. This reference dynamo runs relatively quickly, and is thus useful for extensive explorations. Case E1e-5\_hRm has a similar reversing behaviour to case E1e-4\_hRm.
	Lowering the Ekman number allows for a higher magnetic to kinetic energy ratio ($\overline{M}=3.5$). Dynamo E1e-5\_SF1 has the same Ekman number as E1e-5\_hRm but a ten times smaller Rayleigh number. The magnetic Prandtl number is also slightly higher ($Pm=2$ for E1e-5\_hRm, $Pm=3$ for E1e-5\_SF1). Lowering the Rayleigh number and increasing the magnetic Prandtl number result in a magnetic to kinetic energy ratio larger than 20. The dipolar fraction also significantly increases and the dipole becomes more stable.
	No reversals are observed in this case for the duration of the simulation. The E1e-5\_SF2 is obtained by decreasing again the Rayleigh number by 30\%. This case has the largest magnetic to kinetic energy ratio ($\overline{M}=53$). It is also the reference case with the highest dipolar fraction.
	No reversals are observed in this case. Finally E1e-6\_hRm is obtained using $E=10^{-6}$. 
	The magnetic energy is larger than the kinetic energy, with a similar value on average to case E1e-5\_hRm. No reversals are observed.
	
	Though all the reference cases have a dipole-dominated magnetic field, they lie at a different distance from the dipolar-multipolar transition. \citet{tassin_geomagnetic_2021} found a transition from strong dipoles towards multipolar magnetic field when the ratio of magnetic to kinetic energy becomes lower than about 1. Following this argument, the three high Rm cases (E1e-4\_hRm, E1e-5\_hRm and E1e-6\_hRm) should be close to the transition with a magnetic energy similar or slightly larger than the kinetic energy. Cases E1e-4\_hRm and E1e-5\_hRm are reversing, suggesting indeed that they lie close to the transition. Case E1e-6\_hRm does not reverse polarity for the simulated time, though the eventual occurrence of reversals cannot be excluded due to the limited duration of the simulation.
	Cases E1e-5\_SF1 and E1e-5\_SF2 have magnetic energies significantly higher than the kinetic energies, and can thus be expected to be far from the dipolar-multipolar transition according to the energy ratio criterion.
	
	\subsection{Categories of dynamo behaviours}
	\label{s::behaviors}
	
	Before looking in details into the effect of heat flux heterogeneities, we will first define three distinct dynamical behaviours: (1) ``Dipolar stable'', if $\overline{f_{dip}} >0.25$ (or $\overline{f_{dip}^\dagger} > 0.35$) and no reversals or excursions are observed during the simulation; (2) ``Reversing'', if $\overline{f_{dip}} >0.25$ (or $\overline{f_{dip}^\dagger} > 0.35$) and reversals or excursions are observed during the simulation; (3) ``Multipolar'', if $\overline{f_{dip}} <0.25$ (or $\overline{f_{dip}^\dagger} < 0.35$). The dipolar stable behaviour corresponds to dynamos that are far from a dipolar-multipolar transition. The reversing behaviour corresponds to dynamos that are still dipolar but closer to a transition. 
	In some cases, the dynamo spends a significant amount of time in both the dipolar and the multipolar regime. Using the aforementioned criteria, these bistable dynamos would be categorized as reversing due to their relatively high dipolar fraction on average. A bistable behaviour is however different from the reversing behaviour expected for the Earth, for which we expect the dipolar regime to dominate. We thus define a fourth behaviour: (4) ``Bistable'', if $\overline{f_{dip}}>0.25$ and if $f_{dip}<0.25$ for at least 20\% of the simulation time. 
	
	The separation between the dipolar and multipolar behaviour based on the time-averaged value of $f_{dip}$ is similar to what is done in previous studies \citep{christensen_scaling_2006, soderlund2012influence, tassin_geomagnetic_2021}. A separation between non-reversing and reversing dynamos based on the relative time spent by the dipole below a certain latitude has been previously used \citep{wicht2016gaussian, meduri2021numerical}. Though this kind of separation could work for our simulations, we chose to use our simpler classification that relies only on the statistics of $f_{dip}$. We moreover note that, for bistable dynamos, the bistability appears most clearly in the distribution of $f_{dip}$ while it is much less clear in the distribution of the dipole tilt.
	
\begin{figure}
		\centering
		\includegraphics[width=\linewidth]{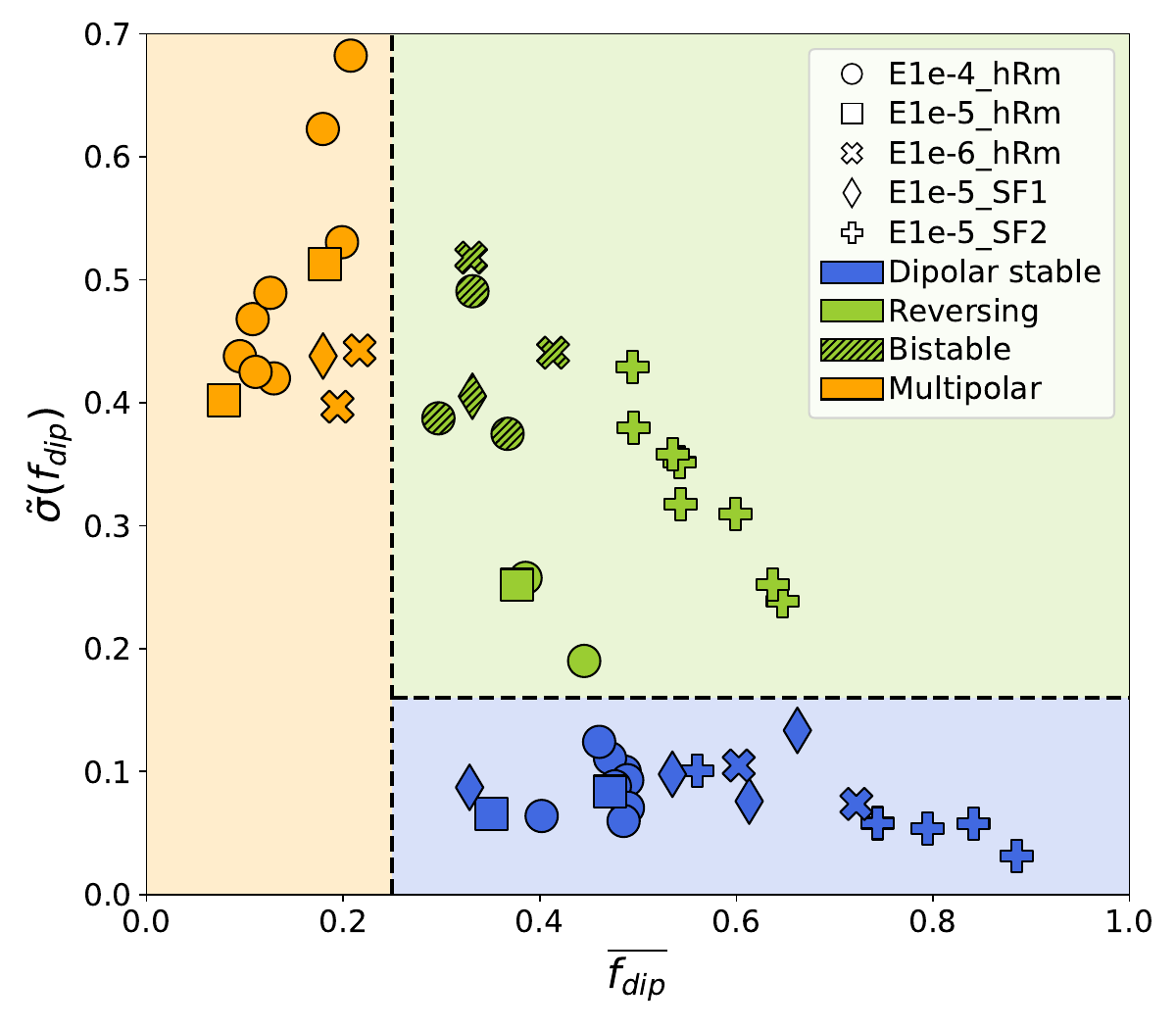}
		\caption{Fluctuation $\tilde{\sigma}(f_{dip}$) and mean value $\overline{f_{dip}}$ of the dipolar fraction at the CMB in all the performed simulations (including simulations with heterogeneous heat fluxes). The marker style shows the reference case from which each dynamo is derived. The colour shows the dynamo behaviour, as defined in section \ref{s::behaviors}.
			All multipolar dynamos plot left of the vertical dashed line at $\overline{f_{dip}}=0.25$, while the horizontal dashed line at $\tilde{\sigma}(f_{dip})=0.16$ separates the dipolar stable and reversing dynamos.}
		\label{mean_fluct}
	\end{figure}

	Depending on the geological period, the behaviour of the Earth's dynamo would be dipolar stable or reversing. With these definitions, the separation between a dipolar and a multipolar behaviour depends directly on the dipolar fraction at the CMB. Although dipolar stable dynamos and reversing dynamos are not discriminated through their dipolar fraction, we can expect the variations of $f_{dip}$ to be higher in the reversing dynamo models, as larger variations of the dipole moment are required for reversals to occur \citep{buffett2013stochastic}. We quantify this variation by defining the relative standard variations of the dipolar fraction as $\tilde{\sigma}(f_{dip}) = \frac{\sigma(f_{dip})}{\overline{f_{dip}}}$ where $\sigma(f_{dip})$ is the standard variation of the dipolar fraction. In Fig. \ref{mean_fluct} are shown the values of $\overline{f_{dip}}$ and $\tilde{\sigma}(f_{dip}$) for all the simulations presented in this study, including simulations with heterogeneous heat fluxes. We indeed find a separation between dipolar stable dynamos and reversing dynamos based on the relative fluctuations of the dipolar fraction. The dipolar stable dynamos all have $\tilde{\sigma}(f_{dip})<0.16$, while all reversing dynamos (and multipolar dynamos) have $\tilde{\sigma}(f_{dip})>0.16$.

	Using these definitions we classify our reference cases E1e-4\_hRm and E1e-5\_hRm as reversing dynamos. The three other cases are classified as dipolar stable dynamos. However, we stress that observing no occurrence of a rare event such as reversals in a finite duration simulation cannot guarantee that it would never happen. 
	
	\subsection{Effect of heat flux heterogeneities for the E1e-4\_hRm case}
	
	In this section, we will explore how the large-scale heat flux patterns presented in Fig. \ref{patterns} affect the behaviour of the dynamo case E1e-4\_hRm. As described in section \ref{s:homogeneous}, this reference case lies close to the dipolar-multipolar transition, showing a reversing behaviour. This reference case can thus be expected to be very sensitive to changes in the flux boundary conditions.
	
\begin{figure*}
		\centering
		\includegraphics[width=1\linewidth]{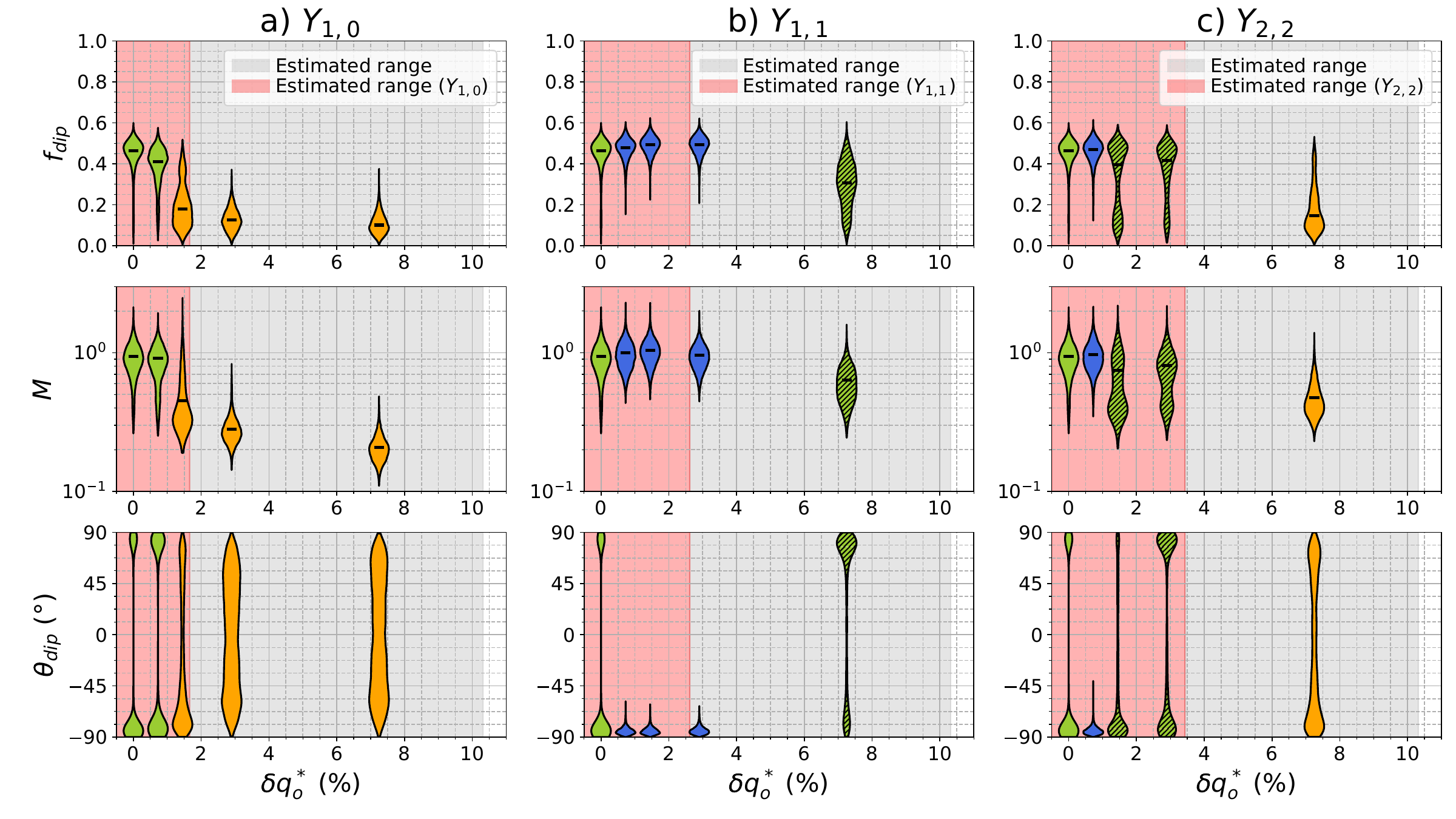}
		\caption{Probability distributions of the dipolar fraction at the CMB ($f_{dip}$), of the ratio of magnetic to kinetic energy ($M$), and of the dipole latitude ($\theta_{dip}$) for the geodynamo simulations derived from the reference case E1e-4\_hRm. The horizontal width of the violins is a measure of the likelihood. The probability distributions are computed using 100 estimation points, each with a relative bandwidth of 0.1. The distributions are given as a function of the amplitude of the heat flux patterns normalized by the inner boundary flux. The colours encode the dynamos category with the same colour code as in Fig. \ref{mean_fluct}. The heat flux patterns are (a) $Y_{1,0}$, (b) $Y_{1,1}$, and (c) $Y_{2,2}$. The expected range of amplitude for the Earth are shown in red for each pattern and in grey for a full pattern (see table \ref{dqo_table}).}
		\label{D1_Y10Y11Y22}
	\end{figure*}

\begin{figure}
		\centering
		\includegraphics[width=\linewidth]{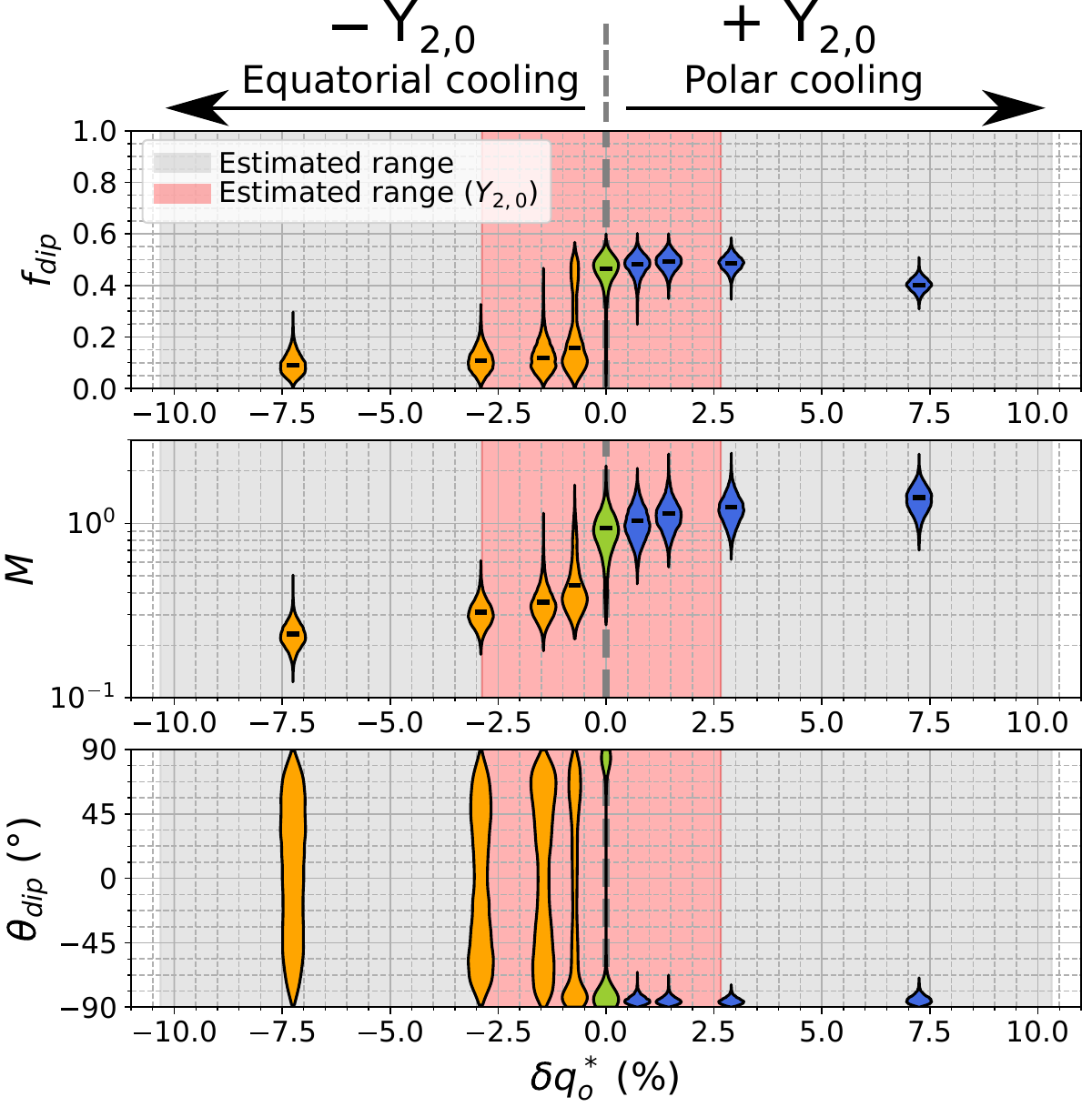}
		\caption{As in Fig. \ref{D1_Y10Y11Y22} for $Y_{2,0}$. Positive amplitudes $\delta q_o^*$ correspond to a polar cooling, while negative amplitudes corresponds to an equatorial cooling.
		}
		\label{D1_Y20}
	\end{figure}

	Violin plots in Fig. \ref{D1_Y10Y11Y22} display the probability distribution of the dipolar fraction, of the ratio of magnetic to kinetic energy, and of the dipole latitude as a function of the pattern amplitude for the  $Y_{1,0}$, $Y_{1,1}$, and $Y_{2,2}$ patterns. The probability distributions for the $\pm Y_{2,0}$ patterns are shown in Fig. \ref{D1_Y20}. The red shaded area shows the expected amplitudes obtained from mantle convection models given in table \ref{dqo_table}. The grey shaded area shows the expected amplitudes for the total heat flux pattern also given in table \ref{dqo_table}. The behaviours described in section \ref{s::behaviors} are well illustrated by the probability distributions. The dipolar stable dynamos have the highest values of $\overline{M}$ and have low fluctuations of both $f_{dip}$ and $M$. They are obtained for moderate amplitudes of the $Y_{1,1}$ pattern, or using the $+Y_{2,0}$ pattern. The dynamo that has the lowest amplitude of the $Y_{2,2}$ pattern also shows a dipolar stable behaviour. The fluctuations of the dipolar fraction in this case are however large, making it very likely for reversals or excursions to eventually occur. The multipolar dynamos show a dipolar fraction that fluctuates around low values, though the dipolar fraction can occasionally reach high values similar to the dipolar dynamos in some cases. The distribution of the dipole latitude covers all the range between -90° and +90°. They are obtained using the $Y_{1,0}$ pattern or the $-Y_{2,0}$ pattern. A multipolar solution is also obtained using a high amplitude of the $Y_{2,2}$ pattern that lies within the expected range for a total heat flux pattern, but that is higher than the expectations for a $Y_{2,2}$ pattern only. The reversing dynamos have similar probability distributions of $f_{dip}$ and $M$ to the dipolar stable dynamos, but with larger tails towards low values. The distribution of the dipole latitude shows two maxima near $\pm 90$°. In some cases, the dynamo spends a significant amount of time in both the dipolar ($f_{dip}>0.25$) and the multipolar ($f_{dip}<0.25$) state. These bistable dynamos are found with the $Y_{2,2}$ pattern for intermediate amplitudes within the estimated range and with the $Y_{1,1}$ pattern for an amplitude larger than the expected range. The bistable nature of these dynamos is well visible in the cases obtained using the $Y_{2,2}$ pattern. In both bistable cases, the probability distribution of the dipolar fraction and the energy ratio is clearly bimodal. The switching between a dipolar state and a multipolar state is visible in the time series of the dipolar fraction as shown in Fig. \ref{bistable}. The bistable dynamo obtained using the $Y_{1,1}$ pattern shows probability distributions that are less clearly bimodal. Nevertheless, this case spent a significant amount of time in a multipolar state because of a relatively low dipolar fraction on average and very large fluctuations. A reversing dynamo is obtained for a weak amplitude of the $Y_{1,0}$ pattern, with a behaviour similar to the reference case,  before the dynamo transition towards a multipolar behaviour for higher amplitudes. 
	
	The opposite effects of the $+Y_{2,0}$ pattern and the $-Y_{2,0}$ pattern is particularly striking. As shown in Fig. \ref{D1_Y20}, the $+Y_{2,0}$ pattern preserves the dipole-dominated structure of the magnetic field, even for large amplitudes, and slightly increases the magnetic energy relatively to the kinetic energy. In contrast, the $-Y_{2,0}$ significantly decreases the energy ratio and triggers a transition towards a multipolar behaviour even for the lowest heterogeneity amplitudes. The increase of the magnetic field intensity or the flow intensity when the $+Y_{2,0}$ or $-Y_{2,0}$ patterns are applied can be seen in more details through the variations of the Elsasser and magnetic Reynolds numbers shown in Fig. \ref{D1_RmEl}.

	\subsection{Systematic study on the effect of equatorial cooling and polar cooling}
	
	The $+Y_{2,0}$ pattern imposes a negative heat flux at low latitudes, and a positive heat flux at higher latitudes. This pattern thus favours a polar cooling of the core, while the equatorial region tends to be stratified. Conversely, the $-Y_{2,0}$ pattern favours an equatorial cooling of the core and acts against convection in higher latitudes. In this section, we will study in more details the effect of both situations on a wide range of dynamo models.
	
	\subsubsection{Effect on high $Rm$ dynamos}
	
	The probability distributions of $f_{dip}$ and $M$ are shown as a function of the amplitude of the $Y_{2,0}$ pattern in Fig. \ref{fdip_EbEu_D1D2D3D4D5}. The geodynamo models E1e-5\_hRm (Fig. \ref{fdip_EbEu_D1D2D3D4D5}b) and E1e-6\_hRm (Fig. \ref{fdip_EbEu_D1D2D3D4D5}c) are affected by the $Y_{2,0}$ pattern in a very similar way to model E1e-4\_hRm (Fig. \ref{fdip_EbEu_D1D2D3D4D5}a). In those three high $Rm$ models, the dynamo becomes multipolar when a $-Y_{2,0}$ heat flux pattern is applied, with a decrease of the magnetic to kinetic energy ratio. For these three cases, the transition occurs for amplitudes of the $-Y_{2,0}$ pattern within the expected range for the Earth. The $+Y_{2,0}$ pattern preserves a strong stable dipole and increases the magnetic energy, even for very large amplitudes. It is noteworthy that very similar effects on the dynamo behaviour are observed for three geodynamo models with Ekman numbers that vary by two orders of magnitude.
	
	\subsubsection{Effect on strong field dynamos}
	
\begin{figure*}
		\centering
		\includegraphics[width=\linewidth]{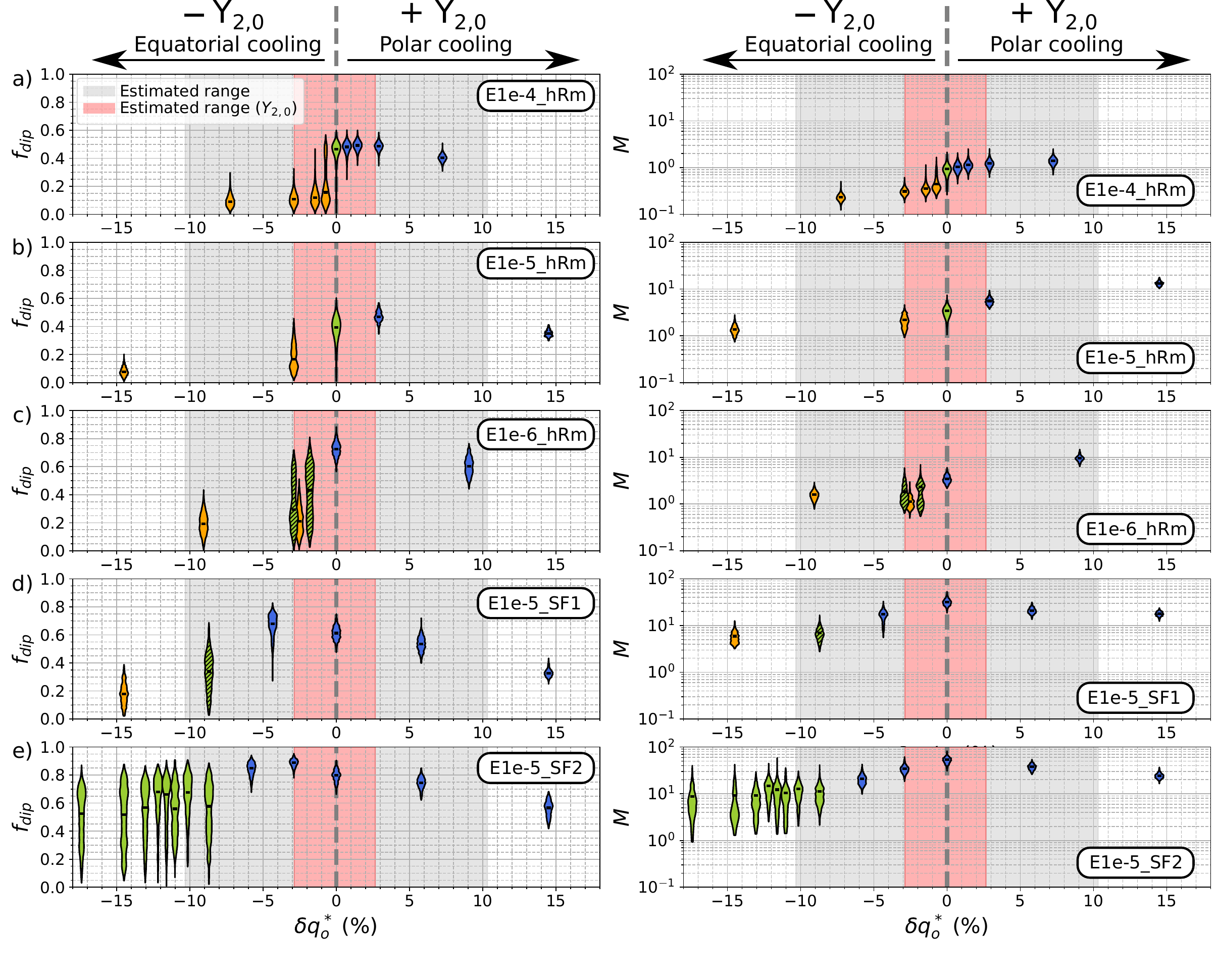}
		\caption{Probability distributions of the dipolar fraction at the CMB ($f_{dip}$) and of the ratio of magnetic to kinetic energy ($M$) for the geodynamo models using the $Y_{2,0}$ pattern. The reference dynamo models are: (a) E1e-4\_hRm, (b) E1e-5\_hRm, (c) E1e-6\_hRm, (d) E1e-5\_SF1, (e) E1e-5\_SF2. The expected range of amplitude for the Earth are shown in red for each pattern and in grey for the full pattern (see table \ref{dqo_table}).}
		\label{fdip_EbEu_D1D2D3D4D5}
	\end{figure*}

	The effect of the $Y_{2,0}$ pattern is very different for the two strong field cases E1e-5\_SF1 (Fig. \ref{fdip_EbEu_D1D2D3D4D5}d) and E1e-5\_SF2 (Fig. \ref{fdip_EbEu_D1D2D3D4D5}e). The amplitude of the magnetic energy relative to the kinetic energy decreases when the amplitude of the pattern is increased, regardless of its sign. The decrease of the magnetic energy is however faster with a $-Y_{2,0}$ pattern. A more prominent difference compared to the high $Rm$ cases exists in the response of the dipolar fraction to the heat flux pattern. For moderate amplitudes ($|\delta q_o^*| \le 5 \%$) the dipolar fraction is higher for a $-Y_{2,0}$ pattern than for a $+Y_{2,0}$ pattern, opposite to the three high $Rm$ cases. For larger amplitudes, the fluctuations of $f_{dip}$ and $M$ significantly increase with the $-Y_{2,0}$ pattern, while the fluctuations do not significantly vary with the $+Y_{2,0}$ pattern. In both the E1e-5\_SF1 case and the E1e-5\_SF2 case, the dynamo behaviour is not significantly affected by the $Y_{2,0}$ pattern for amplitudes within the expected range for the Earth. The dynamo becomes bistable in the E1e-5\_SF1 case and reversing in the E1e-5\_SF2 case for amplitudes of the $-Y_{2,0}$ pattern larger than the expected values for this pattern alone but within the expected range for the total heat flux. The dynamo becomes multipolar in the E1e-5\_SF1 case for larger amplitudes of the $-Y_{2,0}$ pattern that are probably unrealistic for the Earth. No multipolar dynamos are obtained in the E1e-5\_SF2 case, even for unrealistically large amplitudes. In this latter case, the dynamo stays reversing when higher amplitudes of the $-Y_{2,0}$ pattern are used. Interestingly, the transition induced by the $-Y_{2,0}$ pattern from a strongly dipolar dynamo to a multipolar dynamo in E1e-5\_SF1 occurs while the magnetic to kinetic energy ratio stays significantly larger than unity. For example, for the multipolar dynamo obtained with $\delta q_o^* = -14.5\%$, we have $\overline{M}=6$ and $M>2$ for the whole simulation. The value of $\overline{M}$ stays also significantly larger than 1 for the reversing dynamos obtained with the E1e-5\_SF2 model, though the minimum value falls closer to unity in these cases. Such a behaviour is very different from the one observed for the high $Rm$ models, for which the transition to a multipolar state is associated with a decrease of $\overline{M}$ to values that are of order 1 or below.
	
	The eight reversing dynamos in the E1e-5\_SF2 case have fluctuations of the dipolar fraction significantly larger  than the reversing cases obtained with the high $Rm$ models. In this respect, they are more similar to the bistable dynamos obtained in the E1e-5\_SF1 case and in the E1e-4\_hRm case using the $Y_{1,1}$ pattern (see Fig. \ref{D1_Y10Y11Y22}b). The dipolar fraction in these cases can range from almost 0 to 0.9. Similarly, the ratio of magnetic to kinetic energy fluctuates over more than one order of magnitude. They are nevertheless classified as reversing using our criteria as they have a large dipolar fraction on average and they spend a limited amount of time in a multipolar state. In details, the shape of the probability distributions vary between these reversing dynamos. Some of these reversing dynamos spend most of their time in a strongly dipolar state, in which the magnetic energy largely dominates the kinetic energy (see $\delta q_o^*=-12.2\%$ in Fig. \ref{fdip_EbEu_D1D2D3D4D5}(e)). In other cases, the dipolar fraction decreases more frequently and the magnetic energy, while always dominating the kinetic energy, exhibits larger fluctuations (see $\delta q_o^*=-11.0\%$ in Fig. \ref{fdip_EbEu_D1D2D3D4D5}(e)).

	\subsection{Relation between mean zonal flows and the dipole stability}
	
	\subsubsection{Equatorially antisymmetric heat flux ($Y_{1,0}$)}
	
\begin{figure*}
		\centering
		\includegraphics[width=\linewidth]{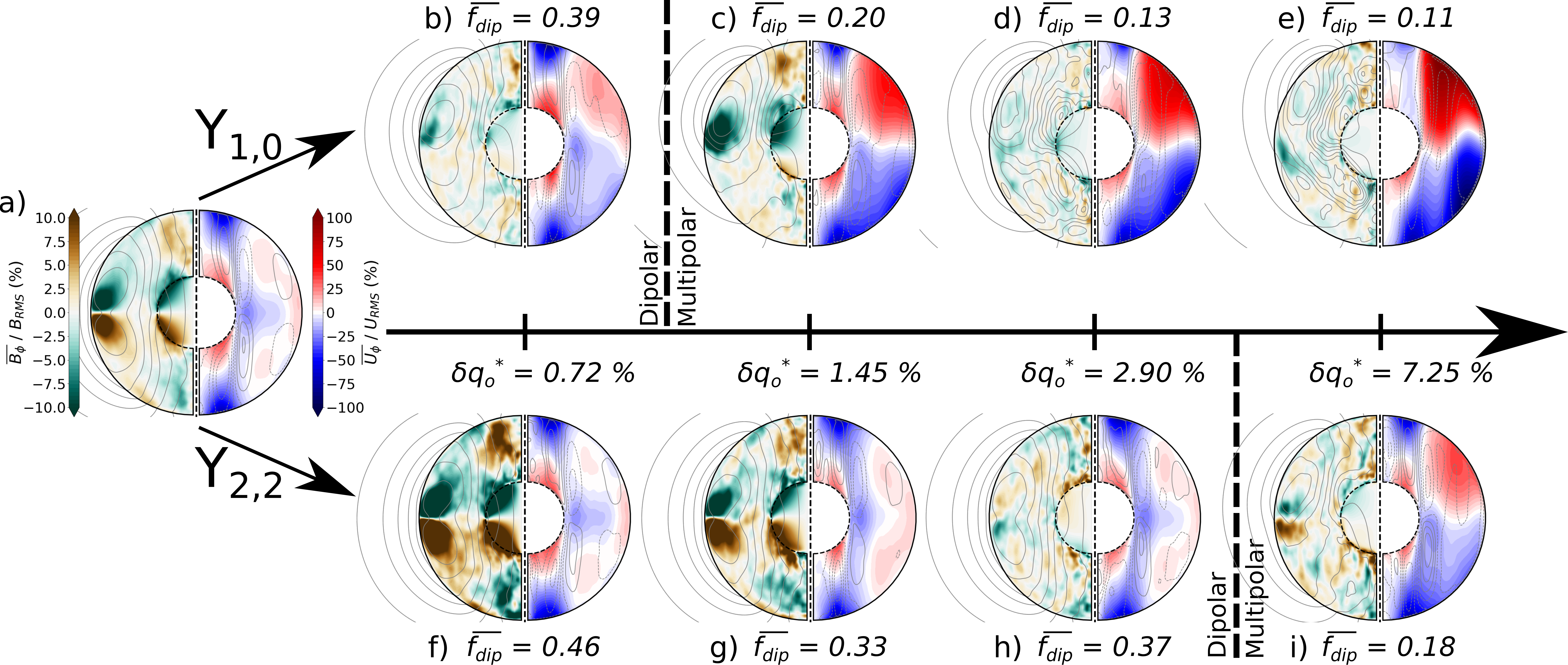}
		\caption{Meridional cuts showing the azimuthal component of the magnetic field and of the velocity averaged in time and in the azimuthal direction relative to the root mean square of the fields for the dynamo cases using the $Y_{1,0}$ and the $Y_{2,2}$ patterns. (a) E1e-4\_hRm reference case without heat flux heterogeneities. (b-e) Dynamo simulations using the $Y_{1,0}$ pattern with an increasing amplitude. (f-i) Dynamo simulations using the $Y_{2,2}$ pattern with an increasing amplitude. For each figure, the azimuthal magnetic field is shown on the left and the azimuthal velocity field is on the right. The grey contours show the poloidal magnetic field lines and the streamlines of the meridional circulation. The flow is clockwise where the field lines are plain and anticlockwise where the field lines are dashed.}
		\label{Y10Y22}
	\end{figure*}

	In Fig. \ref{Y10Y22} we plot the meridional cuts of the azimuthal component of the magnetic field and of the velocity, averaged both in time and in the azimuthal direction, for the simulations using the E1e-4\_hRm model and the $Y_{1,0}$ and $Y_{2,2}$ patterns. The field lines of the poloidal magnetic field and the streamlines of the meridional velocity field averaged in time and in the azimuthal direction are also displayed. In the homogeneous simulation (Fig. \ref{Y10Y22}a), the zonal flow is strong within the cylinder tangent to the inner core (tangent cylinder) and weaker outside. The zonal flow is highly symmetric with respect to the equator and the magnetic field lines show a clear dipolar geometry. The $Y_{1,0}$ pattern breaks the equatorial symmetry in the zonal flow even for the lowest amplitude through the imposed thermal winds \citep[Fig. \ref{Y10Y22}b, see also][]{amit2011influence}.
	The antisymmetry becomes even clearer for larger amplitudes, and the magnetic field becomes multipolar. The $Y_{2,2}$ pattern does not affect significantly the flow for moderate amplitudes, and the magnetic field stays dipolar (Fig. \ref{Y10Y22}f-h). However, for the largest amplitude, the magnetic field becomes multipolar and the zonal flows become strongly antisymmetric (Fig. \ref{Y10Y22}i). Interestingly, zonal flows adopt a geometry very similar to the ones forced by the $Y_{1,0}$ pattern, with an eastward flow in the north of the tangent cylinder and a westward flow in the south.
	
	\subsubsection{Equatorial and polar cooling of the core}
	
\begin{figure*}
		\centering
		\includegraphics[width=\linewidth]{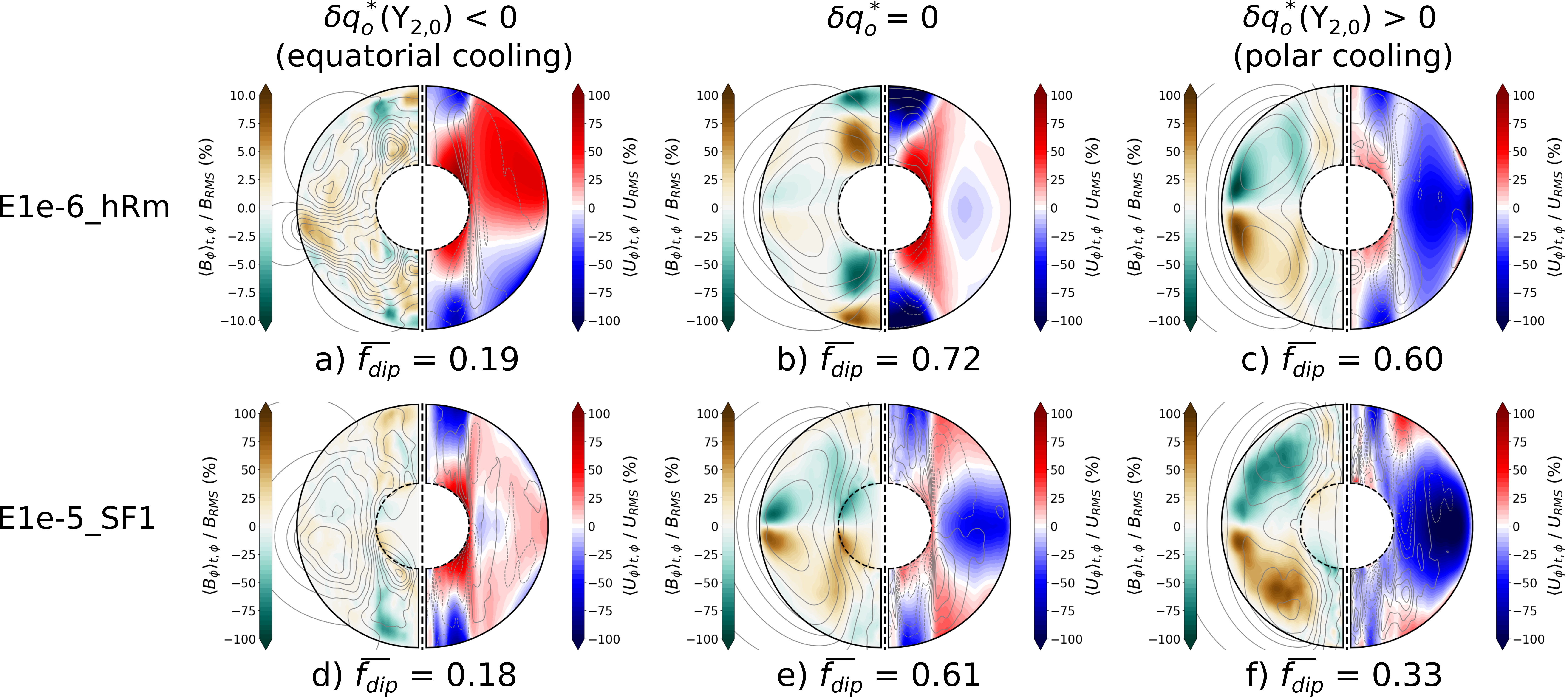}
		\caption{Meridional cuts showing the azimuthal component of the magnetic field and of the velocity averaged in time and in the azimuthal direction relative to the root mean square of the fields for several dynamo cases using a $Y_{2,0}$ heat flux pattern. The time-averaged dipolar fraction of the magnetic field is given in each case. Figures a, b and c correspond to the E1e-6\_hRm case. Figures d, e and f correspond to the E1e-5\_SF1 case. Figures a and d are obtained for a negative $Y_{2,0}$ heat flux pattern (equatorial cooling). Figures c and f are obtained for a positive $Y_{2,0}$ pattern (equatorial cooling). Figures b and e are obtained without heat flux heterogeneities. For each figure, the magnetic field is shown on the left and the velocity field is on the right. The grey contours show the field lines of the poloidal magnetic field and the contours of the meridional circulation. The flow is clockwise where the field lines are plain and anticlockwise where the field lines are dashed. Note that the inner core is insulating for E1e-6\_hRm and conducting for E1e-5\_SF1. The colour scale for the magnetic field in the E1e-6\_hRm case with the $-Y_{2,0}$ pattern is between $\pm10\%$. All the other colour scales are between $\pm100\%$. The heterogeneity amplitudes are: (a) $\delta q_o^* = -9.1\%$, (c) $\delta q_o^* = +9.1\%$, (d) $\delta q_o^* = -14.5 \%$ and (f) $\delta q_o^* = +14.5 \%$.}
		\label{mean_fields}
	\end{figure*}

	We similarly show in Fig. \ref{mean_fields} meridional cuts of the azimuthal velocity and magnetic field for cases E1e-6\_hRm and E1e-5\_SF1 using the $Y_{2,0}$ heat flux pattern. Case E1e-6\_hRm is representative of the three high $Rm$ models, while case E1e-5\_SF1 is representative of the two strong field models. The same meridional cuts for the other dynamo models are shown in Fig. \ref{Y20_full}. Without heat flux heterogeneities, the zonal flow in E1e-6\_hRm (Fig. \ref{mean_fields}b) is similar to the one in E1e-4\_hRm (Fig. \ref{Y10Y22}a).  
	The $+Y_{2,0}$ pattern increases the strength of the westward azimuthal flow outside the tangent cylinder, which becomes of a similar amplitude to the azimuthal flow inside the tangent cylinder (Fig. \ref{mean_fields}c). This results in a larger azimuthal component of the magnetic field outside the tangent cylinder, which can be explained by an increased $\omega$-effect. 
	For an equatorial cooling, the dynamo becomes multipolar and the magnetic field is thus dominated by smaller scales (Fig. \ref{mean_fields}a). The equatorial symmetry of zonal flows is broken outside the tangent cylinder, though the antisymmetry remains lower than what is obtained with the $Y_{1,0}$ pattern and varies depending on the model (see Fig. \ref{Y20_full}). The flow is dominated by a strong eastward flow in the northern hemisphere and a more localized westward flow in the southern hemisphere. The reinforcement of the westward flow for a polar cooling and the occurrence of a mostly eastward flow for an equatorial cooling is consistent with thermal winds driven by the heat flux heterogeneities \citep{dietrich2017reversal}.
	
	The E1e-5\_SF1 case without heat flux heterogeneities (Fig. \ref{mean_fields}e) shows a dipolar magnetic field, similar to the E1e-6\_hRm case. The westward azimuthal flow outside the tangent cylinder is however much stronger in the E1e-5\_SF1 case.
	As in the E1e-6\_hRm case, the $+Y_{2,0}$ pattern strengthens the westward azimuthal flow and the azimuthal component of the magnetic field, while the poloidal magnetic field keeps a dipolar geometry (Fig. \ref{mean_fields}f). As in the E1e-6\_hRm case again, the $-Y_{2,0}$ pattern imposes an eastward azimuthal flow outside the tangent cylinder and the magnetic field becomes multipolar (Fig. \ref{mean_fields}d). Contrary to the high $Rm$ cases, the equatorial symmetry of the zonal flows is maintained. In both high $Rm$ models and strong field models a polar cooling of the core is found to decrease the dipolar fraction of the magnetic field, despite maintaining a non-reversing dipole.
	
\begin{figure}
		\centering
		\includegraphics[width=\linewidth]{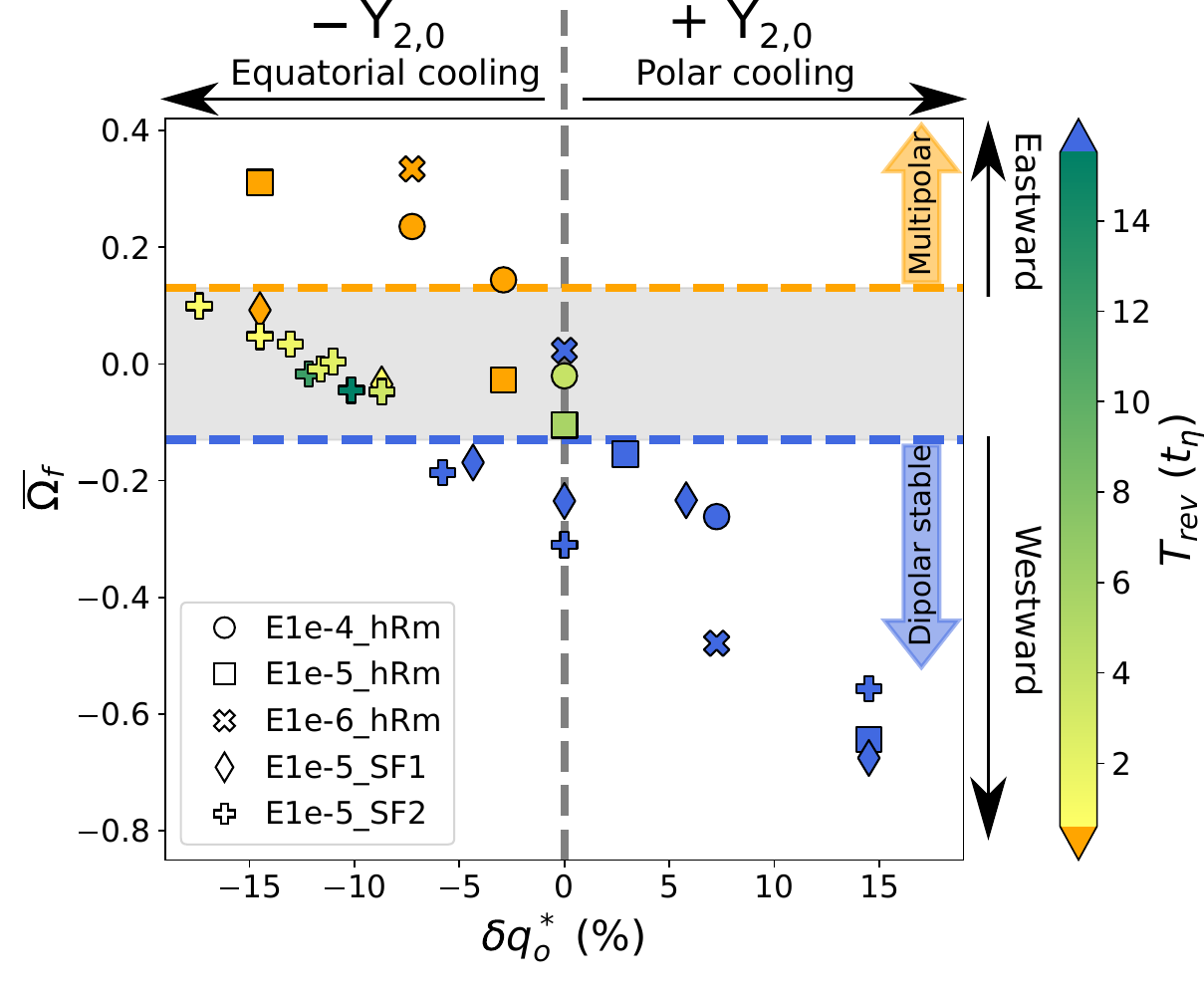}
		\caption{Mean angular velocity of the flow $\overline{\Omega_f}$ as a function of the $Y_{2,0}$ heat flux pattern amplitude. The colour scale shows the average time spent by the magnetic dipole in a single hemisphere (i.e. the average duration of magnetic chrons). The dipolar stable dynamos that do not reverse polarity are shown in blue. The multipolar dynamos showing a weak dipole are shown in orange. Only multipolar dynamos plot above the orange horizontal dashed line at $\overline{\Omega_f}=0.13$, while only dipolar stable dynamos plot below the blue horizontal dashed line at $\overline{\Omega_f}=-0.13$. The gray shaded area highlights the transition between these two regimes, in which dipolar and multipolar dynamos are observed.}
		\label{dqo/Wz}
	\end{figure}

	All the reference dynamos considered in this study have a dominantly westward zonal flow outside the tangent cylinder when averaged in time. The high $Rm$ and the strong field dynamos differ in the strength of this westward zonal flow, the strong field dynamos showing a stronger zonal flow. In order to study the variations of this zonal flow we introduce the mean angular velocity of the fluid relative to the RMS velocity defined as 
	\begin{linenomath}
		\begin{equation}
			\label{Omega_f}
			\Omega_f(t) = \dfrac{L_z(t) \, r_o}{I_0 \, U_{RMS}}
		\end{equation}
	\end{linenomath}
	where $L_z$ is the angular momentum of the liquid core around the spin axis, and $I_0$ is the axial moment of inertia of a homogeneous spherical shell. We chose $\Omega_f$ as a simple proxy for the zonal flows outside the tangent cylinder, as the angular momentum $L_z$ is dominated by flows far from the rotation axis.
	The dependence of $\overline{\Omega_f}$ on the $Y_{2,0}$ pattern amplitude is shown in Fig. \ref{dqo/Wz}. The average time $T_{rev}$ spent by the dipole in a given hemisphere (defined in equation \ref{Trev}) is shown with the colour scale. The dipolar stable dynamos that do not reverse are shown in blue, and the multipolar dynamos are shown in orange. Note that not all the dynamo simulations are shown as the angular momentum has not been recorded for all the simulations (see appendix \ref{app1}). The difference between the amplitudes of the zonal flows between the high $Rm$ and the strong field reference cases is visible by looking at the value of $\overline{\Omega_f}$ for $\delta q_o^*=0$. The mean angular velocity is only a few precents of the RMS velocity in the high $Rm$ cases. $\overline{\Omega_f}$ is negative for both E1e-4\_hRm and E1e-5\_hRm, corresponding to a dominantly westward azimuthal flow. The mean angular velocity is slightly positive for the E1e-6\_hRm case despite the westward zonal flow at the equator shown in Fig. \ref{mean_fields}(b). The strengthening of the westward flows caused by a polar cooling of the core is clearly visible in the decrease of $\overline{\Omega}_f$ for all the models. Only non-reversing dipolar dynamos are obtained for strong westward flows ($\overline{\Omega}_f<-0.13$). Conversely, only multipolar dynamos are associated with strong eastward flows ($\overline{\Omega}_f>0.13$). Weaker zonal flows ($-0.13\le \overline{\Omega}_f \le 0.13$) are associated with both dipolar and multipolar dynamos.
	
	\subsection{Capturing the dipolar-multipolar transition}
\begin{figure*}
		\centering
		\includegraphics[width=\linewidth]{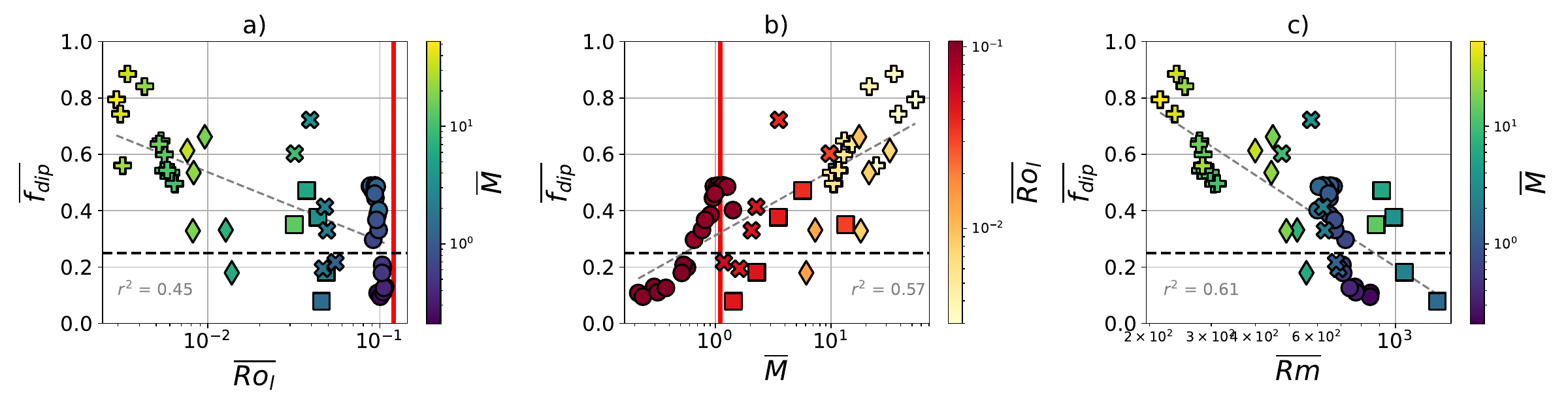}
		\caption{Dipolar fraction as a function of (a) the local Rossby number, (b) the magnetic to kinetic energy ratio, and (c) the magnetic Reynolds number. The horizontal dashed line at a dipolar fraction of 0.25 separates the multipolar dynamos (below) from the dipolar dynamos (above). The transitional value of $Ro_l=0.12$ initially suggested by \citet{christensen_scaling_2006} and of $M=1.1$ suggested by \citet{tassin_geomagnetic_2021} are drawn as the vertical red lines in figures a and b. The markers correspond to the dynamo model used and have the same meaning as in Fig. \ref{mean_fluct}. The colours mark the amplitude of either $\overline{M}$ or $\overline{Ro_l}$, as indicated at the colour bar.
		}
		\label{transitions}
	\end{figure*}

\begin{figure}
		\centering
		\includegraphics[width=\linewidth]{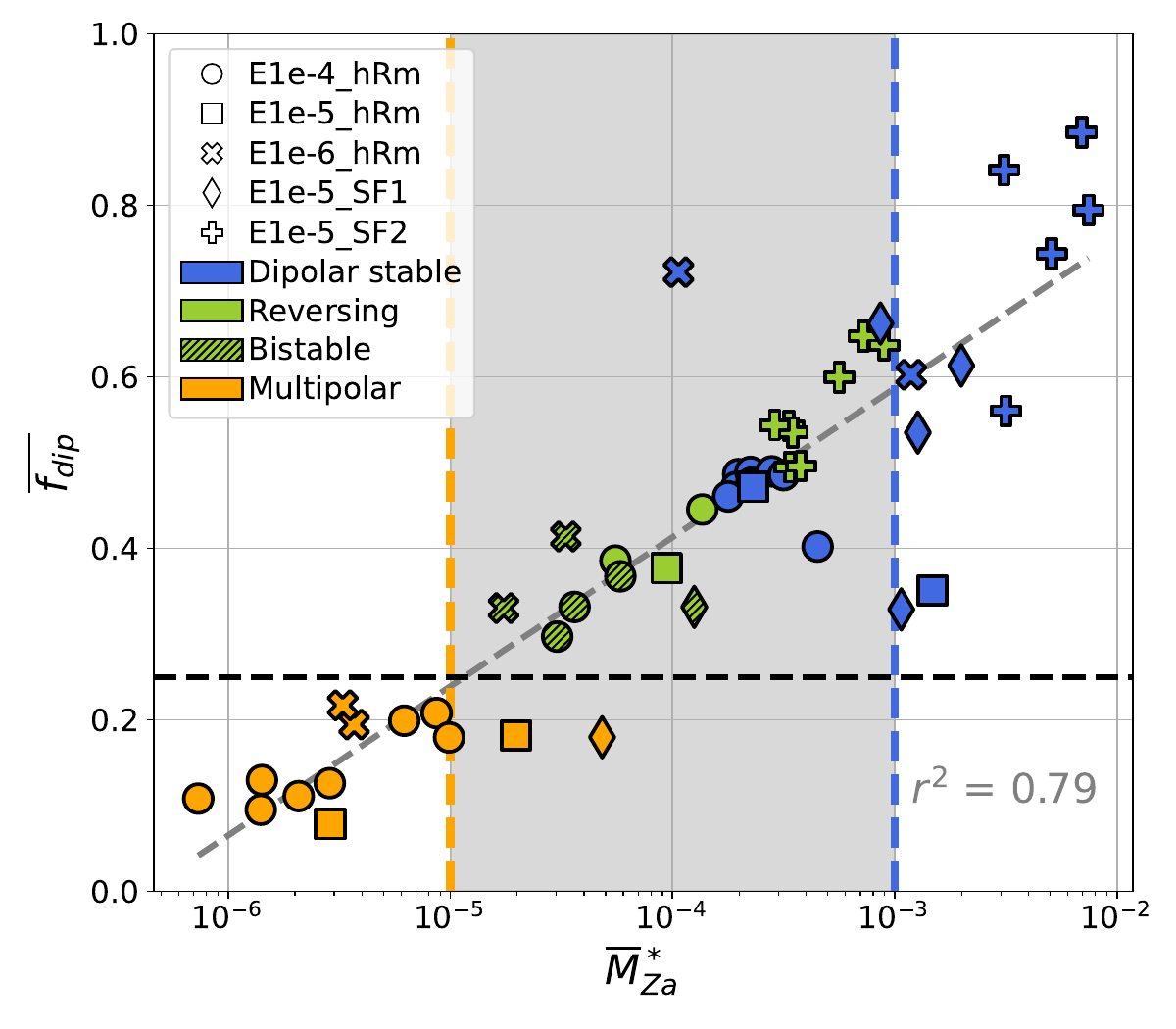}
		\caption{Dipolar fraction of the magnetic field at the CMB as a function of $\overline{M}_{Za}^*$. The marker corresponds to the different reference dynamo models, and the colour gives the behaviour of the dynamo models (same as in figure \ref{mean_fluct}). The orange and blue vertical dashed lines for $M_{Za}^*=10^{-5}$ and $M^*= 10^{-3}$ delimitate the range of $M_{Za}^*$ in which we obtain reversing dynamos.}
		\label{transition2}
	\end{figure}

	Our results show that dipolar dynamos can become multipolar when heterogeneous heat flux patterns are applied at the top of the core. The value of $\delta q_o^*$ for which the  dynamo becomes multipolar depends on the heat flux pattern (see Fig. \ref{D1_Y10Y11Y22} and Fig. \ref{D1_Y20}) and the dynamo models (see Fig. \ref{fdip_EbEu_D1D2D3D4D5}). Several authors suggested that the transition was controlled by the local Rossby number, $Ro_l$ (defined in equation \ref{Rol}), as they observed a transition from dipole-dominated dynamos towards multipolar dynamos when $Ro_l\gtrsim 0.12$ \citep{christensen_scaling_2006, olson_dipole_2006, christensen_dynamo_2010, wicht_theory_2010}. This criteria however fails to capture the transition for strong field dynamos \citep{menu_magnetic_2020,tassin_geomagnetic_2021}, a regime which holds at least for our reference cases E1e-5\_SF1 and E1e-5\_SF2. \citet{tassin_geomagnetic_2021} found that the transition can be better constrained by the ratio between the magnetic and kinetic energy and occurs for $\overline{M}\lesssim1.1$. In Fig. \ref{transitions} we show, for all our dynamo models, the dipolar fraction $\overline{f_{dip}}$ as a function of the local Rossby number, the ratio of magnetic to kinetic energy, and the magnetic Reynolds number. As expected, there is a tendency of a decreased dipolar fraction when $Ro_l$ or $Rm$ is increased and when $M$ is decreased. We quantify this correlation by computing a linear fit between the dipolar fraction and the logarithm of the various criteria. The fit is shown by the grey dashed lines in Fig. \ref{transitions}. This fit shows a poor correlation between the dipolar fraction and the local Rossby number with a correlation coefficient $r^2=0.45$. The correlation is slightly improved with the ratio of magnetic to kinetic energy ($r^2=0.57$) and with the magnetic Reynolds number ($r^2=0.61$). None of these three criteria accurately capture the dipolar-multipolar transition. Both dipolar and multipolar dynamos are obtained for $0.04<\overline{Ro_l}<0.1$ and for $0.6<\overline{M}<5.9$. This shared range is more limited but also exists for the magnetic Reynolds number, with both dipolar and multipolar dynamos found for $558<\overline{Rm}<988$. Note that the transition is well captured by all the three criteria when a single reference dynamo case is considered. The transition nevertheless occurs for different critical values depending on the reference case. All the multipolar dynamos have values of $\overline{Ro_l}$ that are lower than the threshold of $\overline{Ro_l}\simeq0.1$ initially suggested. The transition towards multipolar dynamos occurs at lower values of $\overline{Ro_l}$ when the magnetic to kinetic energy ratio increases. Similarly, the transition towards multipolar dynamos occurs at higher values of $\overline{M}$ when $\overline{Ro_l}$ decreases, yielding multipolar dynamos with $\overline{M}$ as high as 6.
	In Fig. \ref{transition2}, we introduce the parameter $M_{Za}^*$, which is defined as 
	\begin{linenomath}
		\begin{equation}
			\label{def_M*}
			M_{Za}^* = \dfrac{\Lambda_{Za}}{Rm_{Za}^2}
		\end{equation}
	\end{linenomath}
	where $\Lambda_{Za}$ is the Elsasser number based on the zonal antisymmetric component of the magnetic energy and $Rm_{Za}$ is the magnetic Reynolds number based on the zonal antisymmetric component of the kinetic energy. Equivalently, it is also the ratio between the zonal antisymmetric part of the magnetic energy ($E_{{b}{Za}}$) and the zonal antisymmetric part of the kinetic energy ($E_{{u}{Za}}$) multiplied by the magnetic Ekman number $E_{\eta} = \frac{E}{Pm}$, such that $M_{Za}^*=\dfrac{E_{{bZa}}}{E_{{uZa}}}\ E_{\eta}$. In Fig. \ref{transition2} we show $\overline{f_{dip}}$ as a function of $\overline{M}_{Za}^*$ for our set of simulations. We find a better description of the evolution of the dipolar fraction using this parameter, with a correlation coefficient $r^2=0.79$. Only multipolar dynamos are found for $M_{Za}^*<10^{-5}$, while only dipolar stable dynamos are found for $M_{Za}^*>10^{-3}$. In between these two values, dipolar and multipolar dynamos are found. This is also where reversing dynamos are found.
	
	\section{Discussion}
	\label{s:discussion}
	
	\subsection{Effect of heat flux patterns on the stability of the magnetic dipole}
	
	\subsubsection{Destabilizing effect of the $Y_{1,0}$ and $-Y_{2,0}$ patterns}
	
	The magnetic dipole is affected in different ways by the 5 heat flux patterns considered in this study. The $Y_{1,0}$ and $-Y_{2,0}$ heat flux geometries very efficiently destabilize and weaken the magnetic dipole in the E1e-4\_hRm case, while the $+Y_{2,0}$ pattern tends to stabilize the magnetic dipole. The effect obtained for the $Y_{2,0}$ pattern is consistent with previous results that showed a higher reversal frequency when the heat flux is higher at the equator than at the poles \citep{glatzmaier_role_1999,kutzner_simulated_2004,olson_geodynamo_2010}. Note however that \citet{olson_geodynamo_2010} obtained a higher reversal frequency with the $+Y_{2,0}$ pattern than with uniform heat flux conditions, while in our simulations the $+Y_{2,0}$ pattern seems to stabilize the magnetic dipole compared to uniform conditions. We notably confirm that the $-Y_{2,0}$ pattern, which imposes a high heat flux at the equator, is the pattern that affects the most the dynamo behaviour by weakening the magnetic field and destabilizing the magnetic dipole (i.e. it pushes towards a multipolar behaviour). The $Y_{1,0}$ pattern has also been consistently found to moderately destabilize and weaken the magnetic dipole \citep{glatzmaier_role_1999,kutzner_simulated_2004,olson_geodynamo_2010}, as observed in our study.
	
	\subsubsection{Effect of non-zonal heat flux patterns}
	
	\citet{olson_geodynamo_2010} found an increased reversal frequency with the $Y_{1,1}$ pattern. \citet{sahoo_dynamos_2016} on the contrary did not obtain reversals with weak amplitudes of the $Y_{1,1}$ pattern. Our results can reconcile both observations: dipolar stable dynamos are obtained for weak amplitudes of the $Y_{1,1}$ pattern, while a bistable dynamo is obtained for larger amplitudes. The $Y_{2,2}$ pattern has been shown to have a more moderate effect on the dynamo behaviour \citep{kutzner_simulated_2004, olson_geodynamo_2010, sahoo_dynamos_2016}. Nevertheless, \citet{glatzmaier_role_1999} obtain a weaker and less stable dipole with this pattern. We show that the effect of the $Y_{2,2}$ pattern largely depends on its amplitude. We find bistable dynamos for moderate amplitudes, and multipolar dynamos for larger amplitudes. This suggests that the $Y_{2,2}$ pattern slightly destabilizes the dynamo, which becomes multipolar only for a large amplitude of the pattern.
	
	\subsubsection{Comparison between high $Rm$ and strong field models}
	
	The effects discussed above hold for the E1e-4\_hRm case, which is a standard case compared to previous dynamo studies as shown in Fig. \ref{Ra_E_references}. We show in Fig. \ref{fdip_EbEu_D1D2D3D4D5} that the effects of the $\pm Y_{2,0}$ patterns are the same in the two other high $Rm$ cases. The effect however differs in the strong field cases. The $-Y_{2,0}$ pattern is destabilizing in the two strong field dynamos only for very large amplitudes, which are probably unrealistic for the Earth. For more moderate amplitudes, an equatorial cooling of the core strengthens the magnetic dipole. The $+Y_{2,0}$ pattern decreases the dipolar fraction without triggering reversals. \cite{olson_magnetic_2014} showed that an increase in the reversal frequency can be obtained either for an equatorial cooling or for a polar cooling depending on the geodynamo model. They suggest that an equatorial cooling tends to bring the magnetic flux towards the equator, thus destabilizing the axial dipole, while a polar cooling tends to reinforce the underlying convection and thus increase the role of inertia. The first effect, termed as ``geographic control'', would then hold for dynamos close to the onset of dynamo action (low $Rm$), while the second effect, termed as ``inertial control'', would hold for more turbulent dynamos (high $Rm$). We find that an equatorial cooling destabilizes the magnetic dipole in both high $Rm$ models and in strong field models, suggesting that a geographic control could be at play. In strong field dynamos, a polar cooling of the core favours convection as shown by the increase in the magnetic Reynolds number with the $+Y_{2,0}$ pattern in table \ref{table_E1e-5_SF1} for E1e-5\_SF1 and table \ref{table_E1e-5_SF2} for E1e-5\_SF2. However, this strengthening of convection is not associated with a destabilization of the dipole. This lack of inertial control on the dipole stability can be explained by the fact that the strong field models are far from the dipolar-multipolar transition in the dipolar regime, while the models used by \citet{olson_magnetic_2014} lie closer to the transition.

	\subsection{Relation between thermal winds and the magnetic dipole stability}
	\label{wind-dipole}
	
	\subsubsection{Interaction between zonal flows and the magnetic dipole}
	
	We find that the $Y_{2,0}$ pattern affects the zonal flow outside the tangent cylinder through thermal winds, in agreement with previously published models \citep{dietrich2017reversal}: a polar cooling ($+Y_{2,0}$) favours a westward flow outside the tangent cylinder while an equatorial cooling ($-Y_{2,0}$) favours an eastward flow outside the tangent cylinder. 
	Previous studies showed that multipolar dynamo solutions are associated with an increase in the zonal flows \citep{busse2011remarks,gastine2012dipolar,schrinner2012dipole}. This observation can be explained by the role of Maxwell stress: a strong dipolar magnetic field tends to counteract zonal flows and thus decrease the shear that bends the field lines. Conversely, strong zonal flows increase the shear of the poloidal magnetic field, generating an azimuthal component of the magnetic field through an $\omega$-effect \citep{roberts2000geodynamo}.
	We find that a strong polar cooling of the core strengthen the $\omega$-effect, which is associated with a decrease of the dipolar fraction (Fig. \ref{fdip_EbEu_D1D2D3D4D5}). For the two strong field dynamo models, the dipolar fraction is increased for a moderate equatorial cooling. This could be associated in this case with a weakened $\omega$-effect induced by the weakening of the westward flows. 
	
	\subsubsection{Disruption of zonal flows and dipolar-multipolar transition}
	\label{s:mechanisms}
	
	The decrease of the dipolar fraction when increasing the amplitude of the $+Y_{2,0}$ pattern is smooth and is not associated with a destabilization of the magnetic dipole as observed with the $-Y_{2,0}$ pattern. \citet{stanley2010dynamo} showed that a degree 2 order 0 thermal heterogeneity can stabilize or destabilize the dipole if thermal winds generated by the boundary heterogeneity respectively strengthens or counteracts the zonal flows occurring without thermal heterogeneities. This is also what we observe: the dynamo becomes abruptly multipolar under the action of the $-Y_{2,0}$ pattern when the eastward thermal winds outside the tangent cylinder are strong enough to disrupt the otherwise dominantly westward zonal flows. 
	As shown in Fig. \ref{Y10Y22}, the transition imposed by the $Y_{1,0}$ and $Y_{2,2}$ patterns is also associated with a disruption of the westward flows. For these two patterns, the disruption mechanism does not rely on the weakening of westward flows but on the loss of equatorial symmetry in the zonal flows. 
	We find that the $-Y_{2,0}$ pattern also favours antisymmetric flows for the high $Rm$ dynamos, though the antisymmetry is less pronounced in the mean flows than with the $Y_{1,0}$ and $Y_{2,2}$ patterns. 
	Such a loss of equatorial symmetry is expected with the $Y_{1,0}$ pattern that imposes equatorially antisymmetric thermal winds \citep{amit2011influence}. However, the symmetry breaking is spontaneous in the case of the $Y_{2,2}$ and $-Y_{2,0}$ patterns as those patterns do not impose any antisymmetry.
	Solutions not dominated by equatorial symmetry have been previously obtained using the $-Y_{2,0}$ pattern by \citet{cao2014dynamo} for dynamo models driven by a volumetric buoyancy source.
	Without solid inner-core, a spontaneous loss of equatorial symmetry of the mean flow has been observed with homogeneous boundary conditions in dynamo simulations \citep{landeau2011equatorially}, and in double-diffusive convection \citep{monville2019}.
	
\begin{figure}
		\centering
		\includegraphics[width=\linewidth]{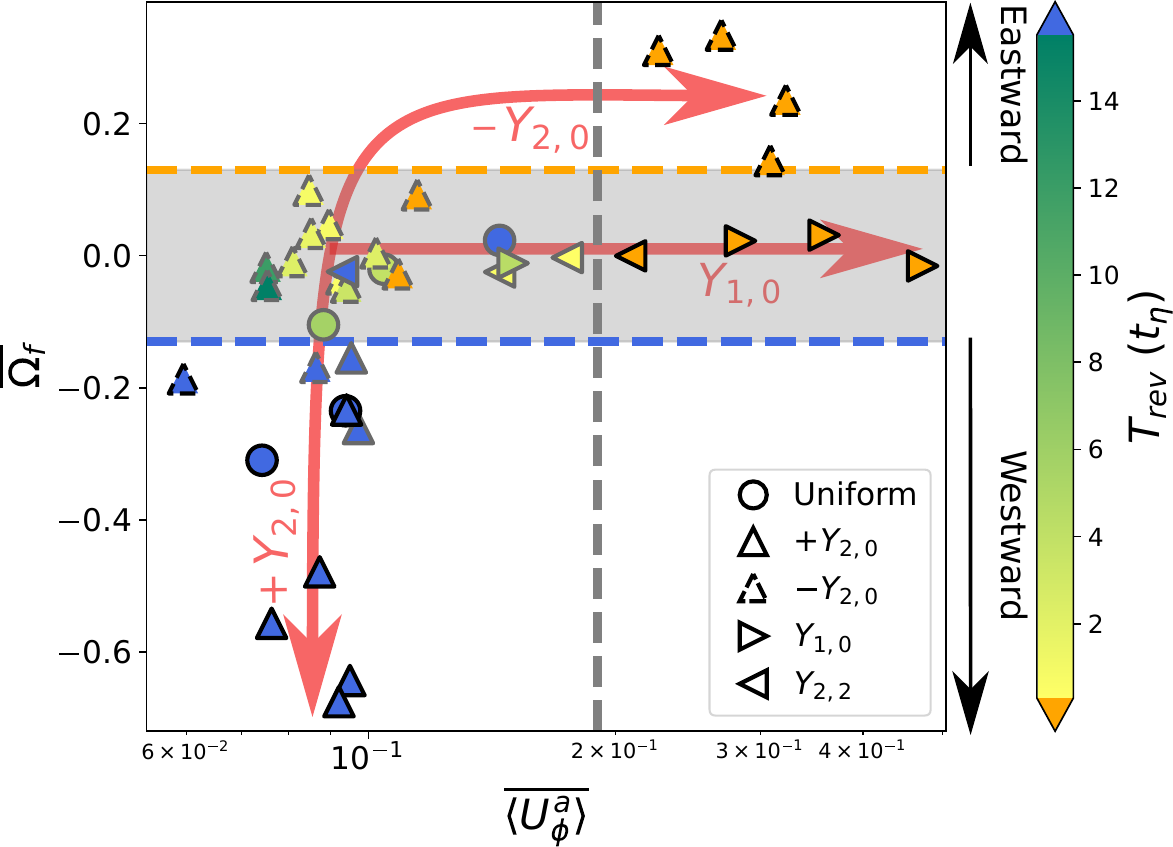}
		\caption{Mean angular velocity of the flow as a function of the RMS antisymmetric zonal flow normalized by the RMS of the full velocity. Dynamos using the $Y_{2,0}$, $Y_{1,0}$ and $Y_{2,2}$ patterns are shown with the upward, rightward and leftward triangles respectively. The reference dynamos with uniform heat flux conditions are shown as circles. The colour scale gives the average reversals period for the reversing dynamos. Dipolar stable dynamos are shown in blue, and multipolar dynamos are shown in orange. Dynamos within the reversing range defined in Fig. \ref{transition2} ($10^{-5}\le M_{Za}^*\le10^{-3}$) are shown with gray contours. The dipolar stable dynamos outside this range in the bottom left of the figure have $M_{Za}^*>10^{-3}$, while the multipolar dynamos outside this range in the top right of the figure have $M_{Za}^*<10^{-5}$. The gray vertical dashed line for $\langle U_{\phi}^a\rangle = 0.19$ separates the range in which only multipolar dynamos are found (on the right-hand side). The orange and blue horizontal dashed lines for $\Omega_f = \pm 0.13$ delimitate the same region, shaded in gray, as the one in Fig. \ref{dqo/Wz}.}
		\label{wz_Uasym}
	\end{figure}

	To quantify the effect of the heat flux pattern on the equatorial symmetry of the flow we define the ratio between the RMS antisymmetric zonal flow and the RMS of the full velocity field, called $\langle U_{\phi}^a \rangle$. In Fig. \ref{wz_Uasym} we illustrate the two mechanisms that can lead to a decreased stability of the magnetic dipole associated with the disruption of the westward flows in our set of simulations. A first mechanism is obtained with the $Y_{1,0}$ pattern (rightward triangles in Fig. \ref{wz_Uasym}) that forcibly breaks the equatorial symmetry of the zonal flows without affecting the space-averaged amplitude of the zonal flow.
	A similar but spontaneous loss of equatorial symmetry is obtained for a large amplitude of the $Y_{2,2}$ pattern (leftward triangles in Fig. \ref{wz_Uasym}), associated with a transition to a multipolar behaviour (see also Fig. \ref{Y10Y22}). A second mechanism is obtained with the $-Y_{2,0}$ pattern (dashed upward triangles in Fig. \ref{wz_Uasym}), which favours dynamos that have weaker westward flows and are reversing or multipolar. In high $Rm$ dynamos, zonal flows become dominantly eastward under the action of the $-Y_{2,0}$ pattern, with a significant loss of equatorial symmetry in most cases. Note that this loss of equatorial symmetry only occurs in the high $Rm$ models, and that the dynamo obtained with the E1e-5\_hRm model and $\delta q_o^*=-2.9\%$ is multipolar without showing a significantly increased equatorial antisymmetric flow (see table \ref{table_E1e-5_SF1}). On the other hand, thermal winds generated by the $+Y_{2,0}$ pattern (plain upward triangles in Fig. \ref{wz_Uasym}) strengthen the equatorially symmetric westward zonal flow, which is associated with non-reversing dipoles.
	
	It has been shown in the context of giant planets that the direction of zonal flows depends on the forcing of convection \citep{gastine2013zonal}. In the geophysical context, westward zonal flows are found to naturally occur in numerical dynamos with homogeneous boundary conditions \citep{kuang1997earth, aubert2005steady, schaeffer_turbulent_2017}. The competition between eastward thermal winds and westward flows could thus explain the widely observed destabilizing effect of the $-Y_{2,0}$ pattern in numerical models \citep{glatzmaier_role_1999, kutzner_simulated_2004, olson_geodynamo_2010, olson_magnetic_2014}.
	According to this mechanism, the $+Y_{2,0}$ pattern should not trigger reversals.
	As mentioned previously, the $+Y_{2,0}$ pattern has been nevertheless found to be destabilizing in some viscosity-dominated geodynamo models \citep{olson_geodynamo_2010, olson_magnetic_2014}.
	While certainly not the only way to destabilize a magnetic dipole, the destabilization mechanism associated with the disruption of the westward flows observed in our low viscosity simulations may be more relevant for the Earth's core.
	
	\subsubsection{Role of equatorial antisymmetry in the dipolar-multipolar transition}
	
	A decrease of the magnetic dipole stability has been found in geodynamo simulations when the equatorially antisymmetric flow is increased \citep{li2002repeated,olson2009dipole,olson_geodynamo_2010,garcia_equatorial_2017} as well as in the VKS experiment \citep{berhanu2010dynamo}. \citet{garcia_equatorial_2017} notably showed that the dipolar-multipolar transition in a set of simulations can be explained by a hydrodynamic transition associated with the loss of equatorial symmetry in the flow. \citet{petrelis2009simple} suggested a reversal mechanism in which breaking the equatorial symmetry of the flow is key. Our results support a role of equatorial symmetry breaking in the dipolar-multipolar transition. Nevertheless, a transition to a multipolar state is obtained without significantly increasing the antisymmetry in the E1e-5\_SF1 model and for moderate amplitudes of the $-Y_{2,0}$ pattern in the E1e-5\_hRm model. Moreover, no systematic increase in the antisymmetric zonal flow is observed in the reversing dynamos. Thus, though the equatorial symmetry breaking seems to be often associated with a destabilization of the magnetic dipole in numerical models and in experiments, it is very likely not the only cause for such destabilization
	
	\subsection{Geophysical interpretation}
	
	\subsubsection{Destabilization mechanisms and westward flows}
	
	Geomagnetic observations support the existence of a dominant eccentric westward gyre, symmetric with respect to the equator, that dominates flow reconstructions at the core-mantle boundary \citep{pais2008quasi, finlay2023gyres}. In our simulations, westward flows naturally occur and are well visible in the time-averaged meridional cuts (see Fig. \ref{Y10Y22} and Fig. \ref{Y20_full}). The westward zonal flow inside the core suggests that destabilizing the present-day dipole would require either an equatorially antisymmetric heat flux ($Y_{1,0}$ mechanism) or a strong equatorial cooling ($-Y_{2,0}$ mechanism). A reinforced polar cooling would on the contrary strengthen the westward flows and stabilize the dipole. The heat flux pattern as deduced from seismic tomography is dominated by an equatorially symmetric $Y_{2,2}$ component \citep{amit_towards_2015}, which does not favour equatorial cooling over polar cooling on average. The effect of such pattern on the amplitude and symmetry of zonal flows needs further studies. Nevertheless, our results  suggest that a tomographic pattern should be only weakly destabilizing for the dynamo.
	
	Interestingly, the implications of these mechanisms can be, to a certain extent, tested for the Earth. The present-day symmetric westward gyre deduced from magnetic observations is compatible with a dipole-dominated dynamo. Magnetic field structures can be reconstructed from paleomagnetic data, and some persistence can be observed in the last thousand to million years \citep{johnson1995time,constable2016persistent}. The recent study by \citet{clizzie2024reversal} notably found that a westward drift of magnetic field structures dominates in the past 100 kyr and that the Laschamp excursion is associated with a transition from eastward drift to westward drift. This persistent westward drift could be a sign of a persistent westward flow, while the change in the drift direction suggests a relation between the dipole stability and changes in the zonal flows.
	
	\subsubsection{Parametrization of the dipolar-multipolar transition}
	
	We show in Fig. \ref{transitions}(a) that the local Rossby number, which was found to control the dipolar-multipolar transition in early dynamo models \citep{olson_dipole_2006, christensen_dynamo_2010}, does not discriminate between strongly dipolar and weakly dipolar dynamos within our simulation set. The inability for this parameter to predict the behaviour of the dynamo was already demonstrated \citep{petitdemange2018systematic,menu_magnetic_2020}. The ratio of magnetic to kinetic energy, put forward by \citet{tassin_geomagnetic_2021}, also fails to describe accurately the transition in our set of simulations as we find multipolar dynamos with a magnetic to kinetic energy ratio as large as 6 (Fig. \ref{transitions}b). The dipolar-multipolar transition imposed by an equatorial cooling in the E1e-5\_SF1 model is particularly interesting in this respect, as it occurs despite the magnetic energy staying significantly larger than the kinetic energy. Our results thus show that heat flux heterogeneities can be a way to destabilize the dipole while keeping a strong magnetic field, as it is expected for the Earth. Though such behaviour is not common in previously published dynamos, other choices in the model's setup can be made to obtain unstable dipoles with large $M$, for example by considering larger Prandtl numbers as demonstrated by \citet{jones2024low}.
	
	We show that the parameter $M_{Za}^*=\dfrac{\Lambda_{Za}}{Rm_{Za}^2}$ (or alternatively $M_{Za}^* = \dfrac{E_{bZa}}{E_{uZa}}\ E_{\eta}$) describes more accurately the evolution of the dipolar fraction. We find a ``reversing range'' for $10^{-5}<\overline{M}_{Za}^*<10^{-3}$ outside of which no reversing dynamos are found. As shown in Fig. \ref{wz_Uasym}, this is within this range that dynamos with weak equatorially symmetric zonal flows are found (corresponding to the markers with grey contours). Dynamos with large $M_{Za}^*$ have strong equatorially symmetric westward flows, while dynamos with low $M_{Za}^*$ have equatorially antisymmetric zonal flows, suggesting that the captured transition is associated with the disruption of westward flows discussed in section \ref{wind-dipole}.
	
	To obtain an estimate of this parameter for the Earth's core, we can write $M \simeq \frac{B_{RMS}^2}{\rho \mu_0 U_{RMS}^2}$. With $B_{RMS}=4$ mT \citep{gillet2010fast}, $U_{RMS}=4\times 10^{-4}$ m s$^{-1}$ \citep{finlay2011flow}, $\rho=1.1\times 10^4$ kg m$^{-3}$ \citep{olson_801_2015}, and $E_{\eta}=10^{-9}$, we have $M\sim7\times 10^3$. We can further assume that the zonal antisymmetric part of the kinetic energy is of about 1\% of the total kinetic energy, consistent with flow reconstructions \citep{fournier2011inference}. Because the magnetic dipole has a zonal antisymmetric geometry and dominates the magnetic energy in the Earth's core, the zonal antisymmetric part of the magnetic energy can be assumed to be not very different from the full magnetic energy. We will here consider that $0.1 \le E_{b_{Za}}/E_b \le 1$. Using these values, we obtain $7\times 10^{-5}\le M_{Za}^*\le7\times 10^{-4}$. This interval lies within the reversing range obtained for our set of simulations, consistent with a  geodynamo that can reverse polarity.
	This suggests that the transition described by the $M_{Za}^*$ parameter could be relevant for the Earth, in contrast with all previously proposed criteria.
	
	The correlation obtained between $\overline{f_{dip}}$ and $\overline{M}_{Za}^*$ can be understood by considering the two terms appearing in the definition of parameter $M_{Za}^*$ in Eq. \ref{def_M*}. The term $\Lambda_{Za}$ mostly quantifies the strength of the magnetic dipole that has a zonal equatorially antisymmetric geometry. It is thus expected to have an increased value of $M_{Za}^*$ when the dipolar fraction is increased. Similarly, $Rm_{Za}$ quantifies the strength of the equatorially antisymmetric zonal flows. Given the relation between the loss of equatorial symmetry and the multipolar transition, one can expect the dipolar fraction to decrease when $Rm_{Za}$ increases.
	It should be noted that a transition from a dipolar state towards a multipolar state is expected to be associated with a decrease of the ratio of magnetic to kinetic energy: the magnetic energy is decreased by the collapse of the dipole, and the kinetic energy can be expected to increase due to a reduced Lorentz force. We remark that all the criteria tested in figure \ref{transitions} suffer from this inherent bias: multipolar dynamos are associated with lower magnetic field and increased velocities, so that combinations that include velocity and/or magnetic field intensity will naturally be discriminant of the transition. The $M_{Za}^*$ parameter nevertheless has the advantage of locating the transition for similar values of $M_{Za}^*$ nearly independently of the considered reference geodynamo model. Nevertheless, all the reference models used in this study share important characteristics such as the mean heat flux, the radius of the inner core or the Prandtl number (fixed to unity in this study). The relevance of $M_{Za}^*$ when varying these parameters should be investigated.
	To make such a vast parametric study tractable, we plan to adopt promising rare event simulation algorithms \citep[e.g.][]{bouchet2019} in future works.
	
	In our simulations, the destabilization of the magnetic dipole due to heat flux heterogeneities is associated with a disruption of the westward flows naturally occurring in the five reference models used in this study. As shown in Fig. \ref{wz_Uasym}, this disruption is caused either by a weakening of the westward flows through thermal winds imposed by an equatorial cooling of the core ($-Y_{2,0}$ pattern), or by a loss of equatorial symmetry in the zonal flows. This loss of equatorial symmetry can be imposed by the $Y_{1,0}$ heat flux pattern, or it can occur spontaneously when increasing the amplitude of the $Y_{2,2}$ or $-Y_{2,0}$ pattern. Conversely, a polar cooling of the core strengthens the symmetric westward flow, and stabilizes the dipole. Our results suggest that the destabilization of the magnetic dipole through heat flux heterogeneities is more difficult in dynamos with strong westward flows. However, we insist that it is not clear from our results whether the westward flow disruption is a cause or a consequence of the multipolar transition. Applied to the Earth's core, this implies that an equatorial cooling of the Earth's core or an equatorially antisymmetric heat flux pattern could disrupt the observed westward zonal flows and destabilize the magnetic dipole, thus increasing the reversal frequency. Conversely, a favoured polar cooling could increase the westward flows and stabilize the dipole.
		
	Because the geodynamo is a highly non-linear system, the effects of combinations of the simple patterns considered here are not obvious.
	We plan to explore the effect of complex heat flux patterns found in mantle convection simulations in a future study.	
	
	As shown in Fig. \ref{transitions}, we find that the local Rossby number and the magnetic to kinetic energy ratio, both previously put forward as controlling the dipolar-multipolar transition, do not capture the dynamo behaviours observed in our simulations. Instead, the dipolar fraction of the magnetic field is found to be correlated with $M_{Za}^*=\dfrac{\Lambda_{Za}}{Rm_{Za}^2}$ (see fig. \ref{transition2}). Dynamos with high $\overline{M}_{Za}^*$ have strong and stable dipoles, while dynamos with low $\overline{M}_{Za}^*$ are multipolar. A transition is found for the range $10^{-5}<\overline{M}_{Za}^*<10^{-3}$ in which all kinds of dynamos are found, including reversing ones.
	Estimates for the Earth fall into this range too, supporting the geophysical relevance of this criterion.
	This parameter is related to the strength and equatorial symmetry of the westward flows, suggesting that geodynamo reversals are associated to the disruption of westward flows.
	According to the relation we obtain between $M_{Za}^*$ and the dynamo behaviour, a superchron-type behaviour would occur for larger values of $M_{Za}^*$. Such increase in $M_{Za}^*$ could arise from a decrease of the anti-symmetric zonal flow component, or from an increase in the magnetic field intensity.
	
	\section*{Data availability}
	The \textit{XSHELLS} code used to run the geodynamo simulations is freely available at\\ \textit{https://nschaeff.bitbucket.io/xshells/}.

	\section*{Acknowledgements}
	
	This research has been supported by the Agence
	Nationale de la Recherche (grant no. ANR-19-CE31-0019). We thank Julien Aubert and Johannes Wicht for thorough reviews on the first version of the manuscript. We also thank Hagay Amit and Alexandre Fournier for their comments. Numerical simulations have been performed using HPC resources of IDRIS, TGCC and CINES under
	allocations A0100407382, A0120407382, A0140407382 and A0160407382 attributed by GENCI (Grand Equipement National de Calcul Intensif). The XSHELLS code received funding from the European High-Performance Computing Joint Undertaking (JU) under Grant Agreement No 101093038 (ChEESE-2P)
	
	\bibliographystyle{gji}
	\bibliography{biblio.bib}
	
	\appendix
	
	\section{Simulation details} 
	\label{app1}
	We give in the following tables some details on the simulations used in this study. The spatial discretisation of the numerical simulation is described in the radial direction by the number of radial points $N_r=N_r^{OC} + N_r^{IC}$ with $N_r^{OC}$ the number of points in the outer core and $N_r^{IC}$ the number of points in the inner core. The scalar fields are expanded in spherical harmonics in the horizontal direction up to the maximum degree $l_{max}$ and the maximum order $m_{max}$. Hyperdiffusion is used in two of the five reference dynamo models on the flow, the magnetic field, and the codensity field. This hyperdiffusion follows the formula suggested by \citet{nataf2015turbulence}. For the flow, the viscosity is given as a function of the spherical harmonic degree $l$ by
	\begin{linenomath}
		\begin{equation}
			\nu(l) = \left\{ \begin{aligned}
				&\nu_0 &\text{ \ for $l\le l_c$}\\
				&\nu_0 q_{\nu}^{l-l_c} &\text{ \ for $l> l_c$}
			\end{aligned}\right.
		\end{equation}
	\end{linenomath}
	with $q_{\nu} = \left(\dfrac{\nu_{max}}{\nu_0}\right)^{\dfrac{1}{l_{max}-l_c}}$.
	Hyperdiffusion is parametrised by the ratio $\frac{\nu_{max}}{\nu_0}$ and the cut-off degree $l_c$, which are given in the tables for the simulations using hyperdiffusion.
	Similarly, the magnetic diffusion is given by 
	\begin{linenomath}
		\begin{equation}
			\eta(l) = \left\{ \begin{aligned}
				&\eta_0 &\text{ \ for $l\le l_c$}\\
				&\eta_0 q_{\eta}^{l-l_c} &\text{ \ for $l> l_c$}
			\end{aligned}\right.
		\end{equation}
	\end{linenomath}
	with $q_{\eta} = \left(\dfrac{\eta_{max}}{\eta_0}\right)^{\dfrac{1}{l_{max}-l_c}}$. Finally, the thermal diffusion follows
	\begin{linenomath}
		\begin{equation}
			\kappa(l) = \left\{ \begin{aligned}
				&\kappa_0 &\text{ \ for $l\le l_c$}\\
				&\kappa_0 q_{\kappa}^{l-l_c} &\text{ \ for $l> l_c$}
			\end{aligned}\right.
		\end{equation}
	\end{linenomath}
	with $q_{\kappa} = \left(\dfrac{\kappa_{max}}{\kappa_0}\right)^{\dfrac{1}{l_{max}-l_c}}$.
	The cut-off degree $l_c$ is the same for the magnetic field and the velocity field.
	
	Table \ref{table_E1e-4_hRm} gives the results for the E1e-4\_hRm model, table \ref{table_E1e-5_hRm} for the E1e-5\_hRm model, table \ref{table_E1e-6_hRm} for the E1e-6\_hRm model, table \ref{table_E1e-5_SF1} for the E1e-5\_SF1 model, and table \ref{table_E1e-5_SF2} for the E1e-5\_SF2 model.

	\begin{table*}
		\centering
		\renewcommand{\arraystretch}{0.8}
		\caption{Characteristics of the simulations using the E1e-4\_hRm reference case with $E=10^{-4}$, $Ra=1.5\times10^8$, $Pm=3$ and $Pr=1$. The spatial discretization is set using $N_r^{OC}=100$, $N_r^{IC}=44$, $l_{max}=100$, and $m_{max}=85$. Hyperdiffusivity has been used for these simulations choosing $\frac{\nu_{max}}{\nu_0}=15$, $\frac{\eta_{max}}{\eta_0}=45$, $\frac{\kappa_{max}}{\kappa_0}=15$ and $l_c=70$. The column $\delta q_o^*$ gives the amplitude of the pattern, $\Delta t_{sim}$ gives the duration of the simulation (after removal of the initial transition time), $\overline{f_{dip}}$ gives the time-averaged value of the dipolar fraction at the CMB, $\tilde{\sigma}\left(f_{dip}\right)$ gives the standard deviation of the dipolar fraction normalized by the mean, $\overline{M}$ gives the time-averaged energy ratio, $M_{Za}^*$ gives the parameter define in equation \ref{def_M*}, $N_{cross}$ gives the number of times the magnetic dipole crosses the equator, $\overline{Rm}$ gives the time-averaged magnetic Reynolds number, $\overline{Ro_l}$ gives the time-averaged local Rossby number, $\overline{\Omega_f}$ gives the time-averaged mean angular velocity, and $\overline{\langle U_{\phi}^a \rangle}$ gives the time-averaged antisymmetric azimuthal part of the flow. The values of $\overline{\Omega_f}$ that have not been computed are marked as ``NC''. The behaviours are DS for dipolar stable, R for reversing, B for bistable and M for multipolar.}
		\setlength{\tabcolsep}{3.2pt}
		\begin{tabular}{ccccccccccccc}
			\toprule
			Pattern  &
			$\delta q_o^*$ & $\Delta t_{sim}$ & $\overline{f_{dip}}$ & 
			$\tilde{\sigma}(f_{dip})$ &  $\overline{M}$ &
			$\overline{M}_{Za}^*$ & $N_{cross}$  & $\overline{Rm}$ & $\overline{Ro_l}$ & $\overline{\Omega_f}$ & $\overline{\langle U_{\phi}^a \rangle}$ & Behaviour\\
			& (\%) & ($t_{\eta}$) & & (\%) & & ($\times 10^{-4}$) &  & & ($\times 10^{-2}$) & (\%) & (\%) & \\ \midrule 
			--- & 0.00 & 134.3 & 0.45 & 19.0 & 0.95 & 1.36 & 34 & 649 & 9.41 & -1.2 & 10.4 & R\\
			Y10 & 0.72 & 35.8 & 0.39 & 25.7 & 0.91 & 0.55 & 8 & 649 & 9.70 & -0.6 & 15.0 & R\\
			Y10 & 1.45 & 34.1 & 0.20 & 53.1 & 0.57 & 0.06 & 148 & 703 & 10.40 & 1.3 & 28.4 & M\\
			Y10 & 2.90 & 33.4 & 0.13 & 42.0 & 0.29 & 0.01 & 285 & 766 & 10.74 & 1.8 & 35.9 & M\\
			Y10 & 7.25 & 33.6 & 0.11 & 46.8 & 0.22 & 0.01 & 256 & 848 & 9.69 & -0.9 & 47.3 & M\\
			Y11 & 0.72 & 38.3 & 0.47 & 11.1 & 1.01 & 1.94 & 0 & 651 & 9.06 & NC & 9.3 & DS\\
			Y11 & 1.45 & 39.3 & 0.49 & 9.3 & 1.05 & 2.24 & 0 & 653 & 8.84 & NC & 9.2 & DS\\
			Y11 & 2.90 & 35.7 & 0.49 & 10.0 & 0.97 & 1.97 & 0 & 667 & 8.74 & NC & 9.6 & DS\\
			Y11 & 7.25 & 39.0 & 0.30 & 38.7 & 0.65 & 0.3 & 81 & 721 & 9.14 & NC & 17.2 & B\\
			Y20 & 0.72 & 37.8 & 0.48 & 8.8 & 1.06 & 2.27 & 0 & 635 & 9.34 & NC & 8.8 & DS\\
			Y20 & 1.45 & 38.0 & 0.49 & 7.1 & 1.17 & 2.79 & 0 & 622 & 9.35 & NC & 8.9 & DS\\
			Y20 & 2.90 & 37.0 & 0.49 & 6.0 & 1.26 & 3.16 & 0 & 605 & 9.58 & NC & 9.2 & DS\\
			Y20 & 7.25 & 35.6 & 0.40 & 6.4 & 1.42 & 4.5 & 0 & 600 & 9.84 & -14.8 & 9.7 & DS\\
			Y20 & -0.72 & 35.5 & 0.21 & 68.2 & 0.53 & 0.09 & 265 & 706 & 10.31 & NC & 23.8 & M\\
			Y20 & -1.45 & 35.2 & 0.13 & 48.9 & 0.37 & 0.03 & 354 & 741 & 10.56 & NC & 28.6 & M\\
			Y20 & -2.90 & 35.7 & 0.11 & 42.5 & 0.32 & 0.02 & 399 & 771 & 10.45 & 8.1 & 30.8 & M\\
			Y20 & -7.25 & 37.4 & 0.10 & 43.8 & 0.24 & 0.01 & 611 & 846 & 10.04 & 13.3 & 32.2 & M\\
			Y22 & 0.72 & 36.3 & 0.46 & 12.4 & 0.99 & 1.78 & 0 & 646 & 9.33 & -1.4 & 9.3 & DS\\
			Y22 & 1.45 & 35.2 & 0.33 & 49.1 & 0.77 & 0.36 & 119 & 674 & 9.83 & -0.2 & 17.5 & B\\
			Y22 & 2.90 & 36.2 & 0.37 & 37.5 & 0.81 & 0.58 & 58 & 670 & 9.54 & -1.4 & 14.4 & B\\
			Y22 & 7.25 & 34.6 & 0.18 & 62.3 & 0.51 & 0.1 & 208 & 710 & 10.28 & -0.0 & 20.9 & M\\
		\end{tabular}
		\label{table_E1e-4_hRm}
	\end{table*}
	
	\begin{table*}
		\centering
		\renewcommand{\arraystretch}{0.8}
		\caption{Characteristics of the simulations using the E1e-5\_hRm reference case with $E=10^{-5}$, $Ra=5\times10^9$, $Pm=2$ and $Pr=1$. The spatial discretization is set using $N_r^{OC}=280$, $N_r^{IC}=80$, $l_{max}=213$, and $m_{max}=170$. The behaviours are DS for dipolar stable, R for reversing, B for bistable and M for multipolar. }
		\setlength{\tabcolsep}{3.2pt}
		\begin{tabular}{ccccccccccccc}
			\toprule
			Pattern  &
			$\delta q_o^*$ & $\Delta t_{sim}$ & $\overline{f_{dip}}$ & 
			$\tilde{\sigma}(f_{dip})$ &  $\overline{M}$ &
			$\overline{M}_{Za}^*$ & $N_{cross}$  & $\overline{Rm}$ & $\overline{Ro_l}$ & $\overline{\Omega_f}$ & $\overline{\langle U_{\phi}^a \rangle}$ & Behaviour\\
			& (\%) & ($t_{\eta}$) & & (\%) & & ($\times 10^{-4}$) &  & & ($\times 10^{-2}$) & (\%) & (\%) & \\ \midrule 
			--- & 0.00 & 96.0 & 0.38 & 25.2 & 3.49 & 0.94 & 17 & 987 & 4.36 & -5.9 & 8.8 & R\\
			Y20 & 2.90 & 1.9 & 0.47 & 8.4 & 5.67 & 2.3 & 0 & 913 & 3.74 & -8.7 & 9.5 & DS\\
			Y20 & 14.50 & 0.8 & 0.35 & 6.6 & 13.3 & 14.78 & 0 & 880 & 3.18 & -36.3 & 9.5 & DS\\
			Y20 & -2.90 & 4.7 & 0.18 & 51.3 & 2.27 & 0.2 & 33 & 1058 & 4.87 & -1.5 & 10.9 & M\\
			Y20 & -14.50 & 5.3 & 0.08 & 40.2 & 1.43 & 0.03 & 125 & 1314 & 4.57 & 17.5 & 22.6 & M\\
		\end{tabular}
		\label{table_E1e-5_hRm}
	\end{table*}
	
	\begin{table*}
		\centering
		\renewcommand{\arraystretch}{0.8}
		\caption{Characteristics of the simulations using the E1e-6\_hRm reference case with $E=10^{-6}$, $Ra=9.6\times10^{11}$, $Pm=0.2$ and $Pr=1$. Hyperdiffusivity has been used for these simulations. The spatial discretization is set using $N_r^{OC}=512$, $N_r^{IC}=0$ (insulating inner core), $l_{max}=341$, and $m_{max}=341$. Hyperdiffusivity has been used for these simulations choosing $\frac{\nu_{max}}{\nu_0}=5$, $\frac{\eta_{max}}{\eta_0}=1$ (no hyperdiffusion on the magnetic field), $\frac{\kappa_{max}}{\eta_0}=5$ and $l_c=307$. The values of $\Omega_f$ that have not been computed are marked as ``NC''. The behaviours are DS for dipolar stable, R for reversing, B for bistable and M for multipolar.}
		\setlength{\tabcolsep}{3.2pt}
		\begin{tabular}{ccccccccccccc}
			\toprule
			Pattern  &
			$\delta q_o^*$ & $\Delta t_{sim}$ & $\overline{f_{dip}}$ & 
			$\tilde{\sigma}(f_{dip})$ &  $\overline{M}$ &
			$\overline{M}_{Za}^*$ & $N_{cross}$  & $\overline{Rm}$ & $\overline{Ro_l}$ & $\overline{\Omega_f}$ & $\overline{\langle U_{\phi}^a \rangle}$ & Behaviour\\
			& (\%) & ($t_{\eta}$) & & (\%) & & ($\times 10^{-4}$) &  & & ($\times 10^{-2}$) & (\%) & (\%) & \\ \midrule 
			--- & 0.00 & 3.7 & 0.72 & 7.4 & 3.53 & 1.06 & 0 & 575 & 3.93 & 1.3 & 14.4 & DS\\
			Y20 & 9.06 & 4.2 & 0.60 & 10.5 & 9.68 & 11.85 & 0 & 476 & 3.20 & -27.0 & 8.7 & DS\\
			Y20 & -1.81 & 15.3 & 0.41 & 44.1 & 2.26 & 0.33 & 51 & 622 & 4.79 & NC & 17.1 & B\\
			Y20 & -2.54 & 5.8 & 0.22 & 44.3 & 1.18 & 0.03 & 74 & 678 & 5.51 & NC & 24.4 & M\\
			Y20 & -2.90 & 10.8 & 0.33 & 51.8 & 2.06 & 0.17 & 74 & 630 & 4.92 & NC & 19.1 & B\\
			Y20 & -9.06 & 8.8 & 0.19 & 39.7 & 1.61 & 0.04 & 135 & 688 & 4.66 & 18.8 & 26.9 & M\\
		\end{tabular}
		\label{table_E1e-6_hRm}
	\end{table*}
	
	\begin{table*}
		\centering
		\renewcommand{\arraystretch}{0.8}
		\caption{Characteristics of the simulations using the E1e-5\_SF1 reference case with $E=10^{-5}$, $Ra=5\times10^8$, $Pm=3$ and $Pr=1$. The spatial discretization is set using $N_r^{OC}=280$, $N_r^{IC}=80$, $l_{max}=213$, and $m_{max}=170$. The behaviours are DS for dipolar stable, R for reversing, B for bistable and M for multipolar.}
		\setlength{\tabcolsep}{3.2pt}
		\begin{tabular}{ccccccccccccc}
			\toprule
			Pattern  &
			$\delta q_o^*$ & $\Delta t_{sim}$ & $\overline{f_{dip}}$ & 
			$\tilde{\sigma}(f_{dip})$ &  $\overline{M}$ &
			$\overline{M}_{Za}^*$ & $N_{cross}$  & $\overline{Rm}$ & $\overline{Ro_l}$ & $\overline{\Omega_f}$ & $\overline{\langle U_{\phi}^a \rangle}$ & Behaviour\\
			& (\%) & ($t_{\eta}$) & & (\%) & & ($\times 10^{-4}$) &  & & ($\times 10^{-2}$) & (\%) & (\%) & \\ \midrule 
			--- & 0.00 & 5.1 & 0.61 & 7.6 & 32.12 & 19.92 & 0 & 399 & 0.76 & -13.3 & 9.4 & DS\\
			Y20 & 5.80 & 3.1 & 0.53 & 9.8 & 21.24 & 12.7 & 0 & 444 & 0.83 & -13.2 & 9.4 & DS\\
			Y20 & 14.50 & 1.8 & 0.33 & 8.7 & 18.03 & 10.7 & 0 & 489 & 0.82 & -38.1 & 9.2 & DS\\
			Y20 & -4.35 & 6.1 & 0.66 & 13.3 & 17.46 & 8.62 & 0 & 448 & 0.96 & -9.5 & 8.6 & DS\\
			Y20 & -8.70 & 9.1 & 0.33 & 40.5 & 7.28 & 1.25 & 15 & 525 & 1.27 & -2.0 & 9.3 & B\\
			Y20 & -14.50 & 2.4 & 0.18 & 43.8 & 6.09 & 0.48 & 17 & 558 & 1.38 & 5.2 & 11.5 & M\\
		\end{tabular}
		\label{table_E1e-5_SF1}
	\end{table*}
	
	\begin{table*}
		\centering
		\renewcommand{\arraystretch}{0.8}
		\caption{Characteristics of the simulations using the E1e-5\_SF2 reference case with $E=10^{-5}$, $Ra=1.5\times10^8$, $Pm=3$ and $Pr=1$. The spatial discretization is set using $N_r^{OC}=280$, $N_r^{IC}=80$, $l_{max}=213$, $m_{max}=170$. The behaviours are DS for dipolar stable, R for reversing, B for bistable and M for multipolar.}
		\setlength{\tabcolsep}{3.2pt}
		\begin{tabular}{ccccccccccccc}
			\toprule
			Pattern  &
			$\delta q_o^*$ & $\Delta t_{sim}$ & $\overline{f_{dip}}$ & 
			$\tilde{\sigma}(f_{dip})$ &  $\overline{M}$ &
			$\overline{M}_{Za}^*$ & $N_{cross}$  & $\overline{Rm}$ & $\overline{Ro_l}$ & $\overline{\Omega_f}$ & $\overline{\langle U_{\phi}^a \rangle}$ & Behaviour\\
			& (\%) & ($t_{\eta}$) & & (\%) & & ($\times 10^{-4}$) &  & & ($\times 10^{-2}$) & (\%) & (\%) & \\ \midrule 
			--- & 0.00 & 5.0 & 0.79 & 5.3 & 53.5 & 74.47 & 0 & 214 & 0.29 & -17.5 & 7.4 & DS\\
			Y20 & 5.80 & 4.3 & 0.74 & 5.8 & 38.01 & 50.91 & 0 & 236 & 0.31 & NC & 7.6 & DS\\
			Y20 & 14.50 & 5.6 & 0.56 & 10.0 & 24.56 & 31.62 & 0 & 283 & 0.32 & -31.4 & 7.6 & DS\\
			Y20 & -2.90 & 7.1 & 0.89 & 3.1 & 34.81 & 69.47 & 0 & 238 & 0.34 & NC & 6.1 & DS\\
			Y20 & -5.80 & 4.8 & 0.84 & 5.7 & 21.33 & 31.07 & 0 & 252 & 0.43 & -10.5 & 6.0 & DS\\
			Y20 & -8.70 & 10.2 & 0.54 & 35.1 & 11.74 & 3.34 & 3 & 279 & 0.55 & -2.7 & 9.4 & R\\
			Y20 & -10.15 & 15.5 & 0.65 & 23.9 & 12.96 & 7.24 & 1 & 278 & 0.53 & -2.5 & 7.5 & R\\
			Y20 & -11.02 & 9.3 & 0.54 & 31.7 & 11.55 & 2.88 & 4 & 287 & 0.58 & 0.2 & 10.2 & R\\
			Y20 & -11.60 & 27.6 & 0.60 & 31.0 & 12.86 & 5.63 & 12 & 284 & 0.56 & -0.5 & 8.1 & R\\
			Y20 & -12.18 & 12.0 & 0.64 & 25.2 & 15.14 & 8.94 & 1 & 276 & 0.52 & -1.0 & 7.5 & R\\
			Y20 & -13.05 & 11.1 & 0.54 & 35.8 & 10.05 & 3.47 & 7 & 298 & 0.61 & 1.9 & 8.5 & R\\
			Y20 & -14.50 & 13.9 & 0.49 & 42.9 & 10.63 & 3.36 & 17 & 304 & 0.64 & 2.7 & 9.0 & R\\
			Y20 & -17.40 & 18.2 & 0.50 & 38.0 & 10.38 & 3.79 & 21 & 311 & 0.64 & 5.6 & 8.5 & R\\
		\end{tabular}
		\label{table_E1e-5_SF2}
	\end{table*}
	
	\FloatBarrier
	
	\section{Dynamo bistability}
	\label{app2}
	
\begin{figure}
		\centering
		\includegraphics[width=\linewidth]{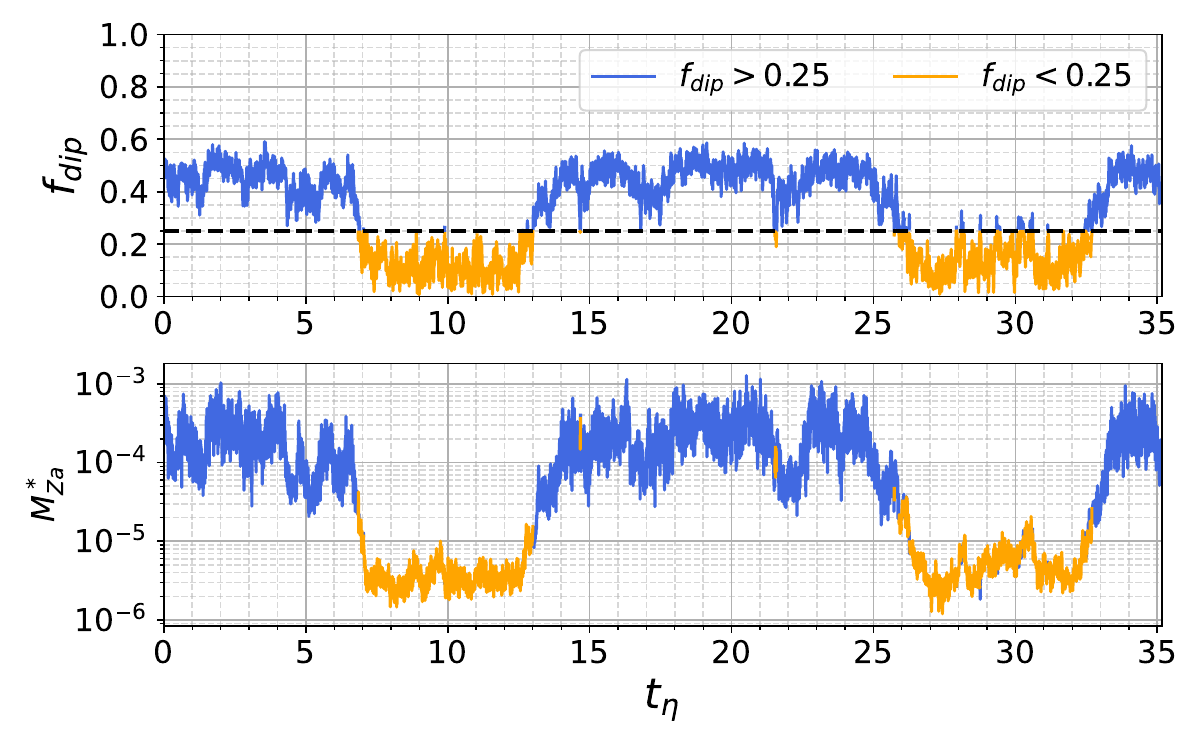}
		\caption{Time serie of the dipolar fraction and of the $M^*$ parameter in a bistable dynamo. The dynamo is obtained using the E1e-4\_hRm reference case and the $Y_{2,2}$ pattern with an amplitude $\delta qo^*=1.4\%$. The time series are shown in blue when the dipolar fraction is higher than 0.25 and in orange when the dipolar fraction is lower than 0.25.}
		\label{bistable}
	\end{figure}

	We classified six of our dynamo simulations as bistable. We recall that we define a dynamo as bistable when $\overline{f_{dip}}>0.25$ and $f_{dip}<0.25$ for more than 20\% of the simulation time. Two of the simulations categorized as bistable have a relatively low dipolar fraction on average with large fluctuations which imply a significant amount of time spent in a multipolar state. The four other bistable cases nonetheless show two clear stable modes and the dynamo switches from one mode to another. In Fig. \ref{bistable} are shown the time series of the dipolar fraction and of $M^*$ for one of the bistable case. This bistable dynamo is obtained using the E1e-4\_hRm reference case and the $Y_{2,2}$ pattern with an amplitude $\delta q_o^* = 1.4\%$ (see Fig. \ref{D1_Y10Y11Y22}). On this figure, the periods when the dynamo is dipolar ($f_{dip}>0.25$) are shown in blue while the periods when the dynamo is multipolar ($f_{dip}<0.25$) are shown in orange. Over the 37 diffusive time simulated, the dynamo switches four times between a dipolar and a multipolar behaviour. The dynamo stays for a long time in both modes, with up to $\sim 13$ $t_{\eta}$ ($\sim$ 520 kyr) spent continuously in a a dipolar mode and $\sim 5$ $t_{\eta}$ ($\sim$ 200 kyr) spent continuously in a multipolar mode. Bistability has been previously documented in dynamo models \citep{simitev2009bistability, busse2011remarks, schrinner2012dipole, petitdemange2018systematic} as well as in the context of the VKS dynamo experiment \citep{berhanu2009bistability}. However, this term usually refers to a hysteresis behaviour in which the dynamo can reach different regimes depending on the initial conditions. Here, we obtain bistable dynamos that oscillate between two significantly different regime in the duration of one single run. 
	
	The bistable dynamos are obtained from originally dipolar simulations when destabilizing heat flux geometries are applied at the top of the core. Importantly, long simulation runs are necessary to identify such a behaviour, as bistable dynamos stay for a long time in either a dipolar or a multipolar regime. As a strongly dipolar magnetic field is used for the initial conditions of all the simulations, shorter simulation runs could have for example falsely led to the interpretation of these bistable cases as dipolar dynamos. 
	
	\section{Effect of heat flux heterogeneities on the magnetic field and flow amplitudes}
	
	\begin{figure}
		\centering
		\includegraphics[width=\linewidth]{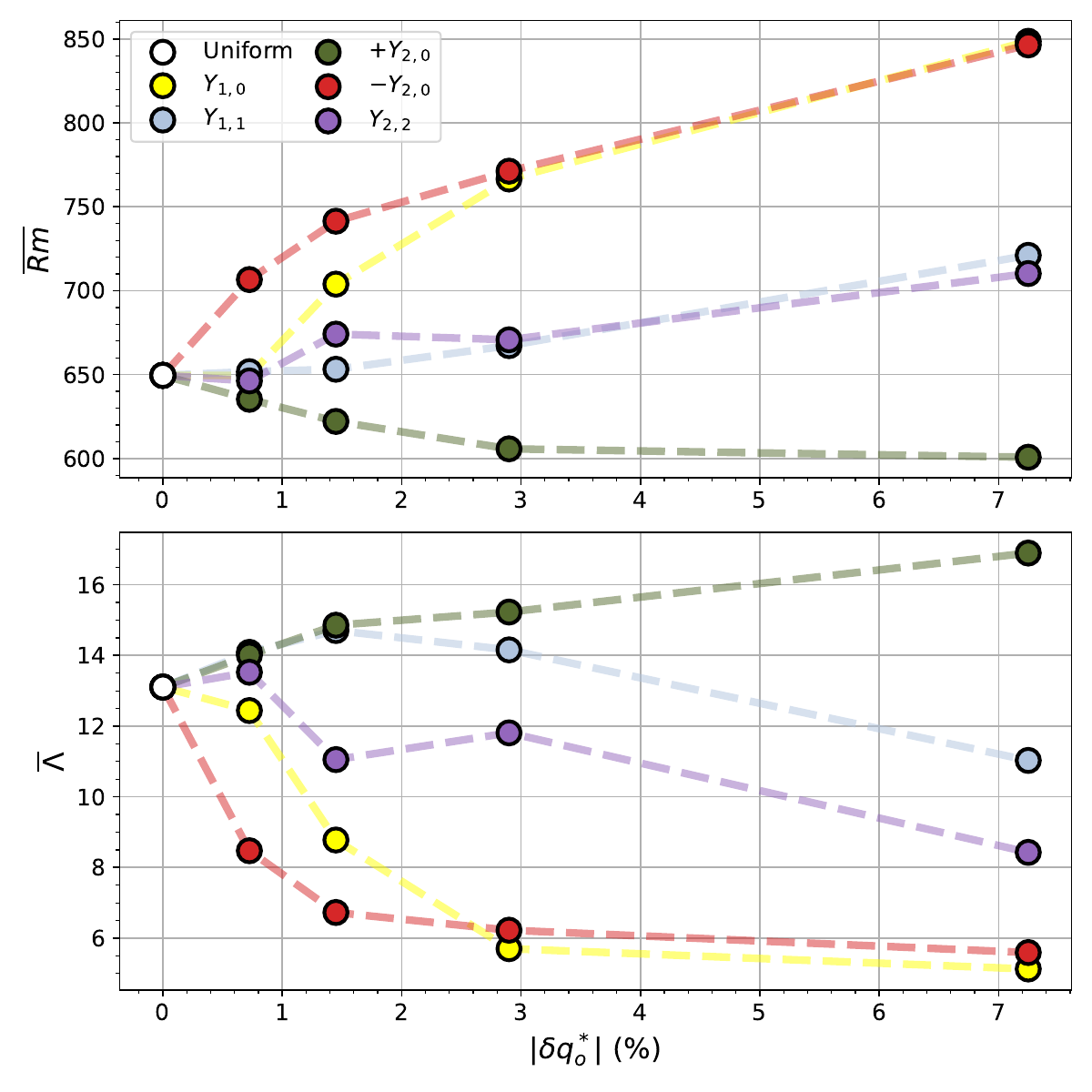}
		\caption{Values of the magnetic Reynolds number (top) and of the Elsasser number (bottom) as a function of the heat flux heterogeneity amplitudes for the different patterns applied on the E1e-4\_hRm reference case.}
		\label{D1_RmEl}
	\end{figure}
	
	The evolutions of the magnetic Reynolds number and the Elsasser number as a function of the pattern amplitudes are shown in Fig. \ref{D1_RmEl}. In the main text, we saw that all the $Y_{1,0}$ and $-Y_{2,0}$ patterns tend to decrease the magnetic to kinetic energy ratio. As it can be seen in Fig. \ref{D1_RmEl}, this translates into both an increase of the strength of the flow, quantified by the magnetic Reynolds number, and a decrease of the magnetic field amplitude, quantified by the Elsasser number. 
	The $-Y_{2,0}$ pattern is the only pattern to induce significant variations in the magnetic Reynolds number and the Elsasser number for the lowest amplitude. The $+Y_{2,0}$ has an opposite effect, by decreasing $Rm$ and increasing $\Lambda$. The opposite effects of the $+Y_{2,0}$ pattern and the $-Y_{2,0}$ pattern is particularly striking. As shown in Fig. \ref{D1_Y20}, the $+Y_{2,0}$ pattern preserves the dipole-dominated structure of the magnetic field, even for large amplitudes, and slightly increases the magnetic energy relatively to the kinetic energy. On the contrary, the $-Y_{2,0}$ significantly decreases the energy ratio and triggers a transition towards a multipolar behaviour even for the lowest heterogeneity amplitudes.
	
	\section{Effect of the $Y_{2,0}$ pattern on the mean flow and magnetic field}
	
\begin{figure*}
		\centering
		\includegraphics[width=\linewidth]{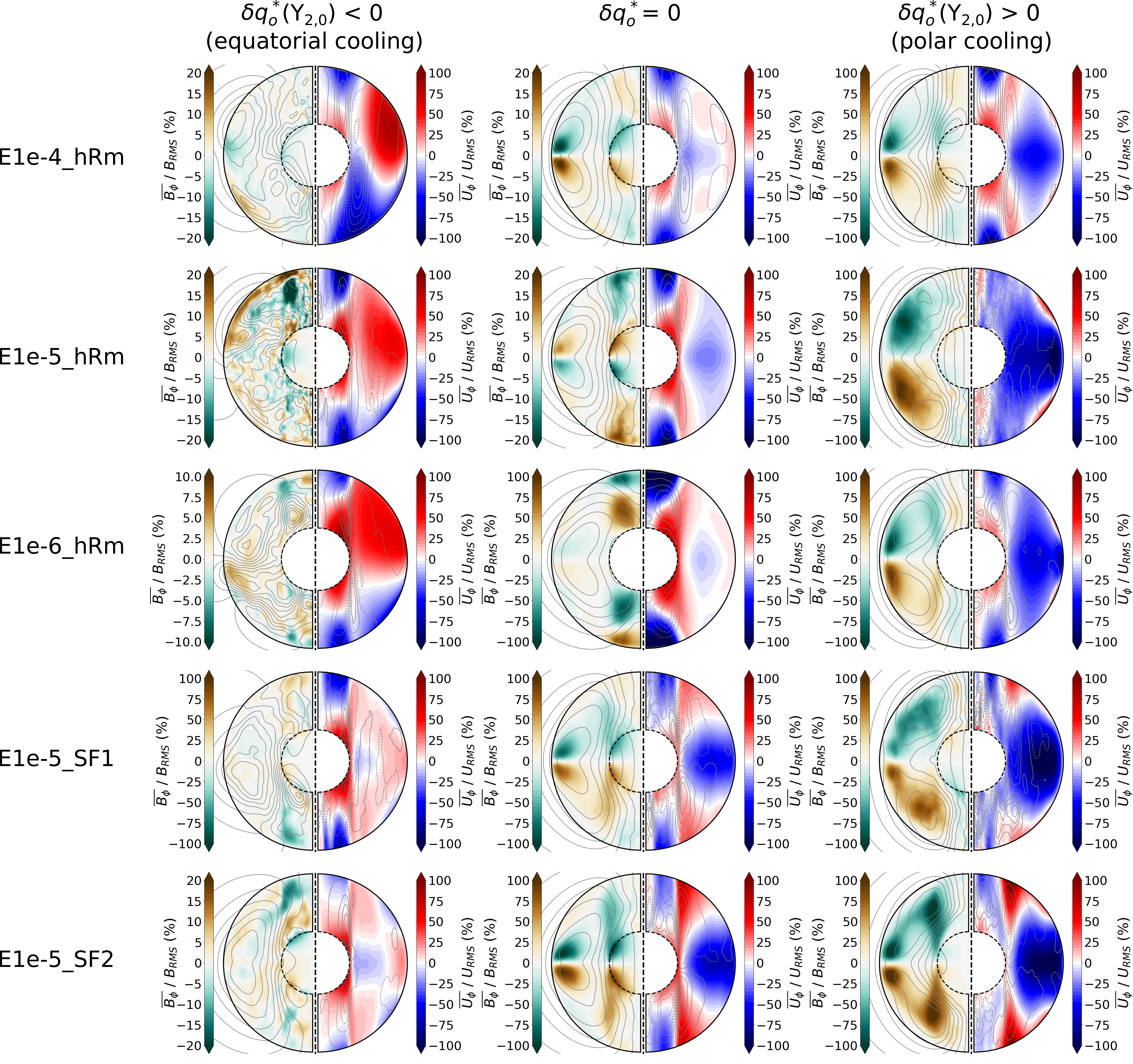}
		\caption{Meridional cuts showing the azimuthal component of the magnetic field and of the velocity averaged in time and in the azimuthal direction relative to the root mean square of the fields for dynamo simulations using the $\pm Y_{2,0}$ pattern. See figure \ref{mean_fields} for more details. The amplitudes of equatorial and polar cooling are given below for each reference dynamo model. E1e-4\_hRm: $\delta q_o^*= \pm 7.2$\%; E1e-5\_hRm: $\delta q_o^*= \pm 14.5$\%; E1e-6\_hRm: $\delta q_o^*= \pm 9.1$\%; E1e-5\_SF1: $\delta q_o^*= \pm 14.5$\%; E1e-5\_SF2: $\delta q_o^*= \pm 14.5$\%. }
		\label{Y20_full}
	\end{figure*}

	We show in Fig. \ref{Y20_full} an extension of Fig. \ref{mean_fields} for the five dynamo models.

	\label{lastpage}

\end{document}